\documentclass[reprint,amsmath,amssymb,aps,longbibliography]{revtex4-1}

\usepackage{graphicx}
\usepackage{dcolumn}
\usepackage{bm}
\usepackage{color}
\usepackage{hyperref}

\begin{document}

\title{Uncovering instabilities in the spatiotemporal dynamics\\of a shear-thickening cornstarch suspension}

\author{Brice Saint-Michel}
\affiliation{Univ Lyon, Ens de Lyon, Univ Claude Bernard, CNRS,
Laboratoire de Physique, F-69342 Lyon, France}%
\altaffiliation{Now at : Department of Chemical Engineering, Imperial College London, London SW7 2AZ, United Kingdom}
  
\author{Thomas Gibaud}%
\email{corresponding author, thomas.gibaud@ens-lyon.fr}
\affiliation{Univ Lyon, Ens de Lyon, Univ Claude Bernard, CNRS,
Laboratoire de Physique, F-69342 Lyon, France}%

\author{S\'{e}bastien Manneville}
\affiliation{Univ Lyon, Ens de Lyon, Univ Claude Bernard, CNRS,
Laboratoire de Physique, F-69342 Lyon, France}%

\date{\today}
\begin{abstract}
Recent theories predict that discontinuous shear-thickening (DST) involves an instability, the nature of which remains elusive. Here, we explore unsteady dynamics in a dense cornstarch suspension by coupling long rheological measurements under constant shear stresses to ultrasound imaging. We demonstrate that unsteadiness in DST results from localized bands that travel along the vorticity direction with a specific signature on the global shear rate response. These propagating events coexist with quiescent phases for stresses slightly above DST onset, resulting in intermittent, turbulent-like dynamics. Deeper into DST, events proliferate, leading to simpler, Gaussian dynamics. We interpret our results in terms of unstable vorticity bands as inferred from recent model and numerical simulations.
\end{abstract}
\pacs{99.XX}
\maketitle

\section{Introduction}

Instabilities are commonly observed in simple, Newtonian fluids when forced to flow under increasingly large Reynolds numbers. Such hydrodynamic instabilities ultimately lead to fully developed turbulence~\cite{frisch1995,Lesieur:2012}, yet following multiple pathways in which vortices~\cite{Busse:1981,Andereck:1986} or turbulent puffs and spots~\cite{Daviaud:1992,Darbyshire:1995} may arise and mediate unsteady or chaotic large-scale flow dynamics. While inertia is at the heart of flow instabilities in simple fluids, non-Newtonian fluids, such as polymer or self-assembled surfactant solutions, may display instabilities at vanishingly small Reynolds numbers due to elasticity~\cite{Larson:1992,Groisman2000} or due to a strong coupling between the flow and the fluid microstructure~\cite{Divoux:2016}. A typical example is provided by wormlike micellar solutions where shear-induced alignment associated with high viscoelasticity leads to shear-banded flows that transition to elastic turbulence at high Weissenberg numbers~\cite{Fardin:2012c}. 

Shear-thickening, the process by which the viscosity of concentrated particulate dispersions dramatically increases above some critical load, is another widespread phenomenon that can be interpreted in terms of an underlying instability. Yet it still remains largely debated and far from fully understood. While the shear-induced growth of particle clusters, referred to as ``hydroclusters'', has long been invoked to explain shear-thickening, especially in Brownian suspensions of small colloidal particles \cite{Brady:1985,Melrose:2004}, it was recently recognized that shear-thickening involves solid friction activated through shear-induced compressive stresses, at least for non-Brownian particles~\cite{Fernandez:2013,seto2013,wyart2014,comtet2017,clavaud2017}. 

In particular, the Wyart and Cates model~\cite{wyart2014} provides a minimal framework where shear-thickening is described as a transition from a low-viscosity, lubricated assembly of nonfrictional particles to a high-viscosity, and possibly fully jammed, frictional contact network. Although supported both by simulations~\cite{Mari2015} and experiments~\cite{pan2015,hermes2016}, this model only provides a zero-dimensional picture that ignores the spatial and dynamical aspects of the transition. In particular, the prediction of hysteresis associated with S-shaped flow curves hints at an instability that should involve the separation of the system into shear bands oriented along the vorticity direction of the flow and referred to as vorticity bands in the literature~\cite{Olmsted:2008}. In the specific case of hard non-Brownian particles, stress balance across the interface between bands prevents the existence of steady vorticity bands~\cite{hermes2016,chacko2018}. This could explain the large temporal fluctuations that are ubiquitously observed in shear-thickening systems~\cite{lootens2003,Brown:2014,pan2015,hermes2016,Bossis2017}. 

The exact nature of the instability and its origin remain however elusive. For instance, it is unclear whether particle migration, free surface instability, vorticity shear-banding and/or gradient shear-banding are at play~\cite{fall2015,Nagahiro:2016,rathee2017} and whether the instability involves truly chaotic dynamics~\cite{hermes2016,grob2016} as contained in early phenomenological models~\cite{Fielding2004,Aradian2005}. Therefore a full spatiotemporal picture of such unstable flows is required from experiments that rely on sufficiently long and statistically stationary measurements.  

Here we focus on the dynamics of a non-Brownian dispersion of cornstarch particles that displays discontinuous shear-thickening (DST) in the hysteretic region of the phase diagram~\cite{wyart2014}. Dense starch suspensions are a popular system to investigate shear-thickening~\cite{fall2008,brown2009,wagner2009,waitukaitis2012,crawford2013,Peters2016,Han2016}, even though they raise technical challenges linked to particle polydispersity and porosity, to sedimentation and potential migration under shear, or to possible adhesion between particles~\cite{fall2015,Han:2017,chatte2018}. Using a rheometer in a concentric-cylinder geometry, we solicit the suspension at a constant shear stress and simultaneously image the 
local flow behavior at a ``mesoscopic'' scale of a few particle sizes with ultrasonic echography. We first provide a detailed statistical analysis of the shear rate fluctuations as a function of the distance to the DST onset. These global dynamics show intermittent, turbulent-like statistics close to DST onset that progressively give way to Gaussian statistics far above DST. We then turn to the spatially and temporally resolved features of the flow to get deeper physical insights in such dynamical regimes. Our major findings are (i) that the bulk suspension remains homogeneously sheared whatever the applied stress in the DST regime and (ii) that the dynamics of the shear rate at short time scales display a diffusive behavior scattered with ballistic events which correspond to unstable macroscopic bands that propagate along the vorticity direction and proliferate as the stress is increased. 

\section{Statistical analysis of the shear rate dynamics}
\label{s:analysis}

\begin{figure}
	\centering
    \includegraphics[scale = 0.53]{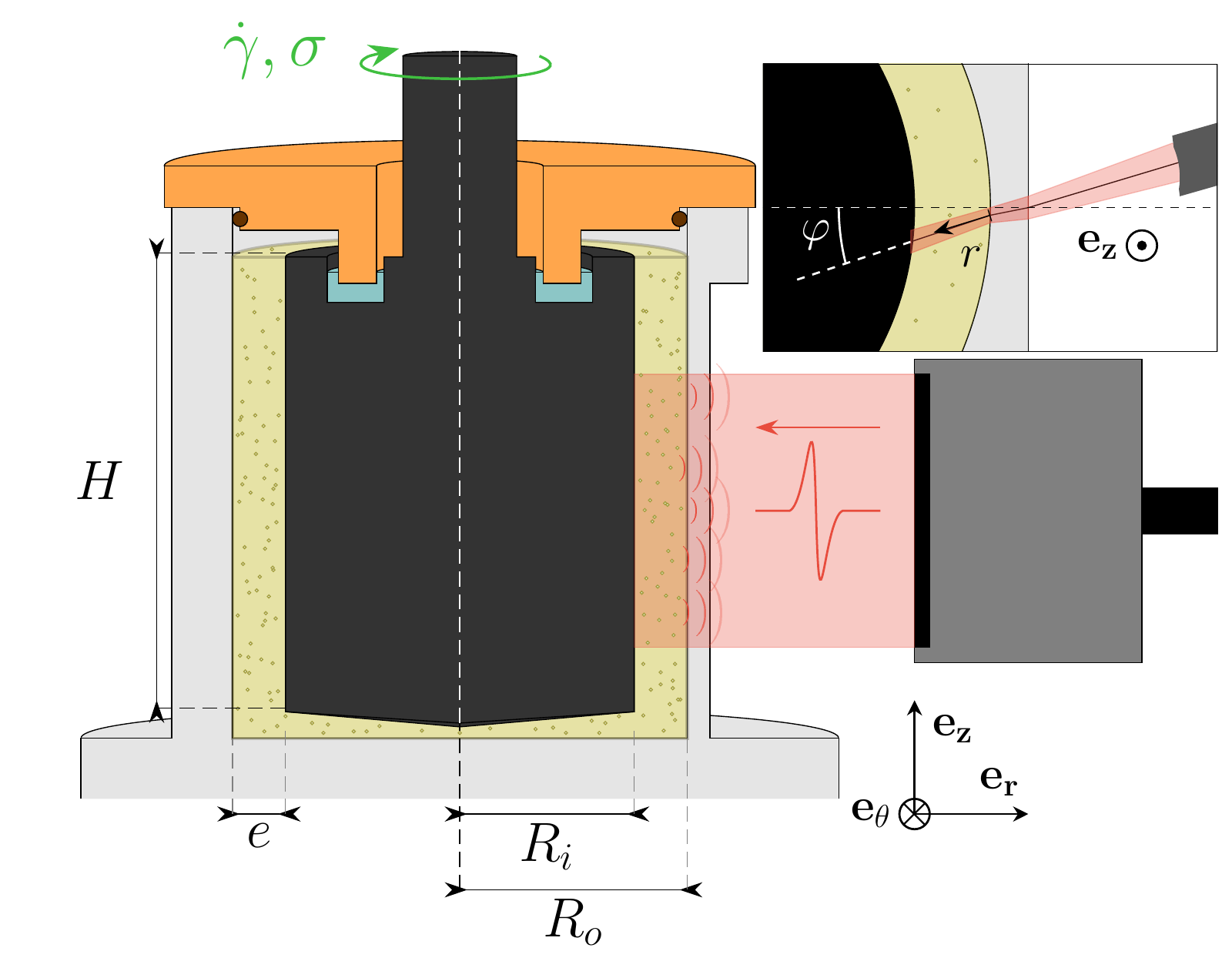}
    \caption{Vertical cross-section of the experimental setup used to study shear-thickening in a dense cornstarch suspension under cylindrical Taylor-Couette flow. The suspension (in yellow) fills the gap of a concentric-cylinder device of height $H=63$~mm and width $e=R_o-R_i=2$~mm, where $R_o=25$~mm is the diameter of the fixed outer cylinder (in gray) and $R_i=23$~mm is the diameter of the rotating inner cylinder (in black) attached to a rheometer that imposes a given stress $\sigma$ and measures the corresponding shear rate $\dot\gamma$. The lid used to prevent evaporation is shown in orange. An ultrasonic probe (right) is used to image the flowing suspension. The top view in the inset shows the ultrasonic beam crossing the gap at an angle $\varphi\simeq 5^\circ$ that allows one to map the local tangential velocity of the suspension. See also Fig.~\ref{fig:setup_pics} in Appendix~\ref{s:rheo} for pictures of the setup.}
	\label{fig:setup_sketch}
\end{figure}

\begin{figure*}
	\centering
    \includegraphics[scale = 0.9]{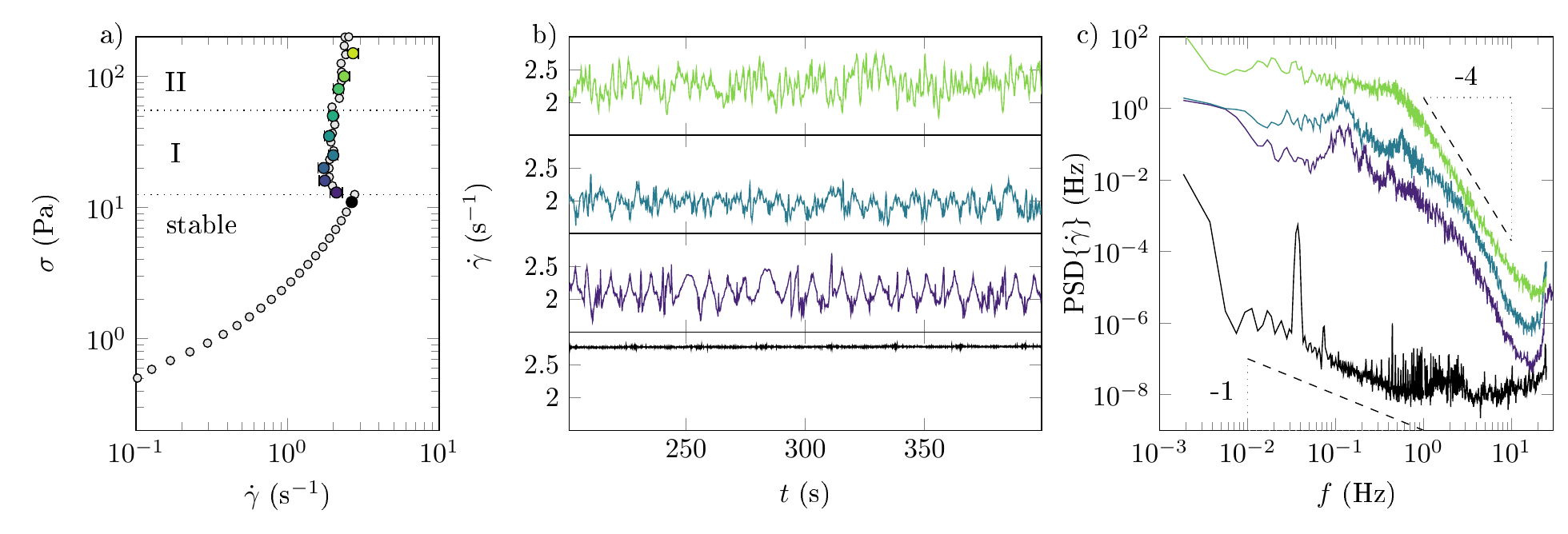}
    \caption{Mechanical response to shear stress of a dense cornstarch suspension at 41\% wt. (a) Flow curve recorded in a small-gap concentric-cylinder geometry by ramping up the imposed stress $\sigma$ with 40~s per point (grey circles) and through subsequent long experiments of at least 2,500~s each under constant stresses (colored circles, the shear rate $\dot\gamma$ is averaged over 2500~s and error bars correspond to the standard deviation). The dashed lines indicate the limits of the various DST regimes discussed in the text.
(b) Snapshots of long time series of $\dot\gamma (t)$ for four imposed stresses : $\sigma=11$ (black), $13$ (purple), $25$ (blue) and $100$~Pa (green) from bottom to top.
(c) Power spectral density of $\dot\gamma(t)$ for the four imposed stresses shown in (b) with the same colors. For clarity, the PSDs at $25$ and $100$~Pa were shifted vertically by a factor 10 and 100 respectively.}
	\label{fig:fc_tseries_spectra}
\end{figure*}

The mechanical response of a cornstarch dispersion at 41\% wt. in a density-matched mixture of water and cesium chloride (Appendix~\ref{s:sample}) is monitored under a cylindrical Taylor-Couette flow generated by a stress-controlled rheometer (TA Instruments ARG2). A small-gap concentric-cylinder geometry, sketched in Fig.~\ref{fig:setup_sketch}, was carefully designed to minimize solvent evaporation, inertia, particle sedimentation and migration, and to avoid instability of the free surface and bubbles trapped in the dispersion, allowing for measurements over at least 20 hours (Appendix~\ref{s:rheo}). Upon an increasing ramp of imposed shear stress, the flow curve, shear stress $\sigma$ {\it vs} shear rate $\dot\gamma$, of our cornstarch dispersion displays the hallmark of DST. As seen in Fig.~\ref{fig:fc_tseries_spectra}a, the shear rate increases smoothly with $\sigma$ up to a critical stress $\sigma_c\simeq 12$~Pa at which $\dot\gamma$ shows a sudden discontinuity. Above $\sigma_c$, $\dot\gamma$ remains on average constant and equal to $\dot\gamma_c\simeq 2$~s$^{-1}$ i.e. the viscosity $\eta=\sigma/\dot\gamma_c$ increases linearly with stress, which is a signature of DST~\cite{Brown:2014}. As already reported in previous studies~\cite{pan2015,hermes2016}, the shear rate exhibits complex fluctuations over the whole vertical part of the flow curve for $\sigma>\sigma_c$. 

\begin{figure*}
	\centering
    \includegraphics[scale = 0.95]{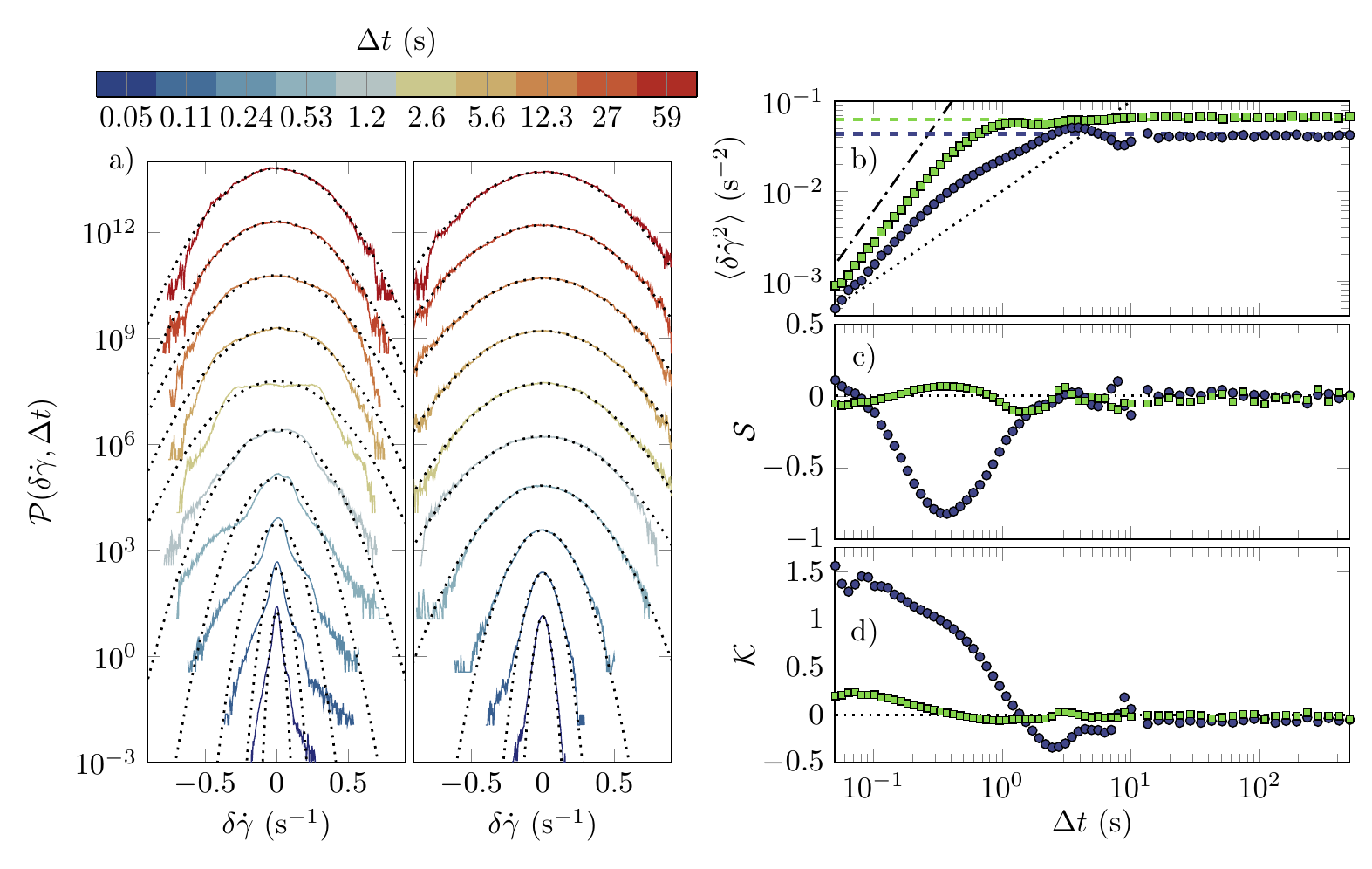}
    \caption{Statistical properties of the shear rate increments $\delta\dot\gamma(t,\Delta t)$ across the DST transition. (a) Probability density functions $
{\mathcal{P}}(\delta\dot\gamma,\Delta t)$ of the increments of $\dot\gamma(t)$ in regime I (left, $\sigma=16$~Pa) and in regime II (right, $\sigma=100$~Pa) for several time lags $\Delta t$ (see color scale). Dotted lines show Gaussian distributions with the same variances as $\delta\dot\gamma(t,\Delta t)$. For clarity, successive PDFs were shifted vertically by $10^{3/2}$ starting from $\Delta t=0.05$~s. (b) Variance, (c) skewness and (d) kurtosis of the shear rate increments for $\sigma=16$~Pa (blue circles) and $100$~Pa (green squares). In (b) the thick dashed lines correspond to twice the variance of the shear rate, $2(\langle \dot\gamma(t)^2\rangle_t - \langle \dot\gamma(t)\rangle_t^2)$, and the thin dotted and dash-dotted lines indicate $\langle\delta\dot\gamma^2\rangle\propto\Delta t$ and $\Delta t^2$ respectively.}
	\label{fig:pdfinc_structfun}
\end{figure*}

To gain insight into these dynamics, we impose various constant shear stresses ranging from 5 to 200~Pa during long steps of at least 2,500~s each. Portions of the resulting $\dot\gamma(t)$ time series are shown over 200~s in Fig.~\ref{fig:fc_tseries_spectra}b, see also Fig.~\ref{fig:Maizena41_recap_20170922_28} in Appendix~\ref{s:stat} for the full data set. The fluctuations of $\dot\gamma$ are statistically stationary and the average values coincide with the flow curve measured through a stress sweep (compare colored and grey symbols in Fig.~\ref{fig:fc_tseries_spectra}a).

The power spectral density (PSD) of the shear rate is displayed in Fig.~\ref{fig:fc_tseries_spectra}c. We may distinguish three different regimes depending on the imposed stress $\sigma$. Below $\sigma_c$, fluctuations of $\dot\gamma$ are negligible and the PSD simply reflects the sum of mechanical and electrical noises: it roughly decreases as $1/f$ at low frequencies before reaching noise level (black curve in Fig.~\ref{fig:fc_tseries_spectra}c). The large peak at $f\simeq 0.04$~Hz corresponds to the rotation period of the inner cylinder, which always shows in the PSDs below $\sigma_c$ due to slight imperfections of the apparatus. For $\sigma_c < \sigma \lesssim 55$~Pa, the PSD shows a complex series of peaks at low frequencies followed by a strong decrease at higher frequencies (purple and blue curves in Fig.~\ref{fig:fc_tseries_spectra}c). For even higher stresses $\sigma \gtrsim 55$~Pa, the peaks give way to a plateau with a cutoff frequency of about 0.5~Hz above which the spectrum decreases as $f^{-4}$. Thus, based on the shape of the PSDs, we hereafter identify a stable regime for $\sigma<\sigma_c$, a first unstable and statistically complex DST regime for $\sigma_c < \sigma \lesssim 55$~Pa (regime I), and a second unstable yet statistically simpler DST regime for $55 \lesssim \sigma \lesssim 200$~Pa (regime II). Note that we do not claim that regime II represents the ultimate stage of the DST transition as instability of the free surface above 200~Pa prevented higher stress values to be explored while ensuring homogeneous shearing conditions.

To further analyze the fine properties of the shear rate fluctuations in both regimes I and II, we now borrow statistical tools from turbulence physics \cite{castaing1990,chevillard2012} and focus our attention on the statistical properties of the {\it increments} of the shear rate $\delta\dot\gamma (t,\Delta t) = \dot\gamma(t + \Delta t) - \dot\gamma(t)$. For a given value of the time lag $\Delta t$, we first introduce the probability density function (PDF) $
{\mathcal{P}}(\delta\dot\gamma,\Delta t)$ of $\delta\dot\gamma(t,\Delta t)$ taken as a random variable in time, see Fig.~\ref{fig:pdfinc_structfun}a. We then investigate the second, third and fourth moments of $\delta\dot\gamma$, respectively through the variance $\langle\delta\dot\gamma^2\rangle$, the skewness ${\mathcal{S}}$ and the kurtosis ${\mathcal{K}}$ displayed in Fig.~\ref{fig:pdfinc_structfun}b--d as a function of $\Delta t$ (see Appendix~\ref{s:stat} for definitions).

In regime II, ${\mathcal{P}}(\delta\dot\gamma,\Delta t)$ remains Gaussian for all time lags $\Delta t$ whereas in regime I, it shows strong asymmetry with exponential tails at short time scales and only assumes a Gaussian shape at long time scales ($\Delta t\gtrsim 10$~s in the specific case of $\sigma=16$~Pa shown in Fig.~\ref{fig:pdfinc_structfun}). Such a striking difference between regimes I and II shows even more clearly on the moments of the increments. For both regimes I and II, $\langle \delta\dot\gamma^2\rangle$ saturates to twice the variance of the shear rate, which indicates statistical independence of $\dot\gamma(t)$ and $\dot\gamma(t+\Delta t)$ at large $\Delta t$. Meanwhile ${\mathcal{S}}$ and ${\mathcal{K}}$ both converge to zero, confirming  loss of correlation and Gaussian statistics for large time lags. At short time lags, however, the dynamics appear much less trivial and differ significantly. In regime II, the skewness and the kurtosis are featureless and always close to zero, consistently with Gaussian distributions. Still, the variance displays a {\it ballistic} growth, i.e. $\langle \delta\dot\gamma^2\rangle\propto \Delta t^2$, before plateauing to its asymptotic value for $\Delta t\gtrsim 1$~s. In regime I, a similar initial ballistic-like behavior is followed by {\it diffusive} dynamics, i.e. $\langle \delta\dot\gamma^2\rangle\propto \Delta t$, for $0.5\lesssim\Delta t\lesssim 5$~s, coupled to a strong negative peak in the skewness and large decreasing values of the kurtosis. Together with exponential tails in the PDFs, this suggests the presence of intermittent events with large negative amplitude at short time scales in regime I.

To summarize, above $\sigma_c$, the shear rate first displays non-Gaussian, intermittent fluctuations at short time scales, characterized by a cross-over between ballistic and diffusive dynamics as the time lag increases (regime I). As the shear stress is increased deeper into DST, the fluctuations become fully Gaussian with only a ballistic behavior at short time scales (regime II). In order to unveil the origin of such complex dynamics, we turn to ultrasonic imaging which allows us to map the local tangential velocity $v(r,z,t)$ of the cornstarch suspension as a function of the radial direction $r$ across the gap, the vorticity direction $z$ and time $t$ \cite{gallot2013} (Appendix~\ref{s:echo}).

\section{Local insights from ultrasonic imaging}

\begin{figure*}
	\centering
    \includegraphics[scale = 1]{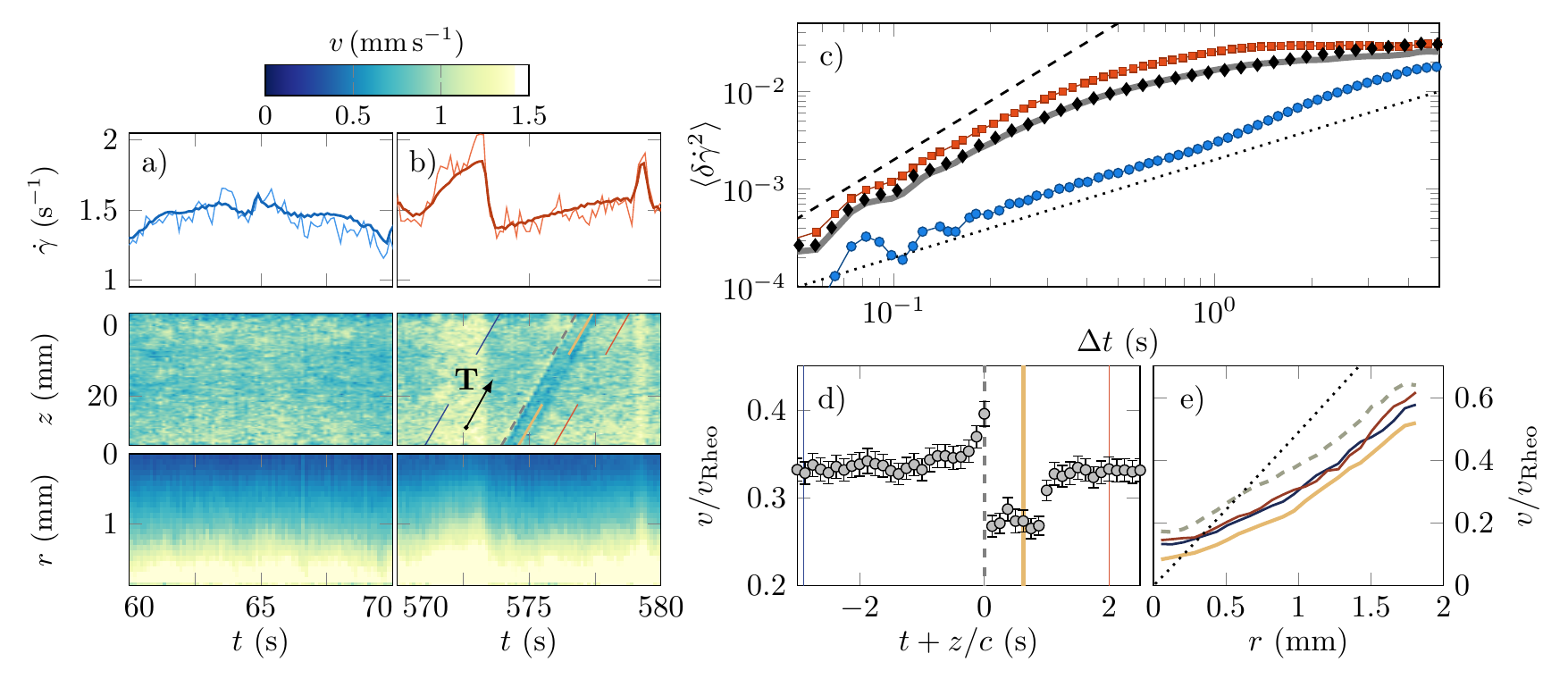}
    \caption{Spatiotemporal patterns governing the dynamics in regime I ($\sigma = 12$~Pa). (a)~Quiesent phase and (b)~propagating event. From top to bottom: global shear rate $\dot\gamma(t)$ (thick line) and effective shear rate in the bulk material $\dot\gamma_B(t)$ inferred from ultrasonic imaging and rescaled by a multiplicative factor 2.55 (thin line); spatiotemporal maps of the local velocity $v(r,z,t)$ averaged in the $r$- and $z$-directions respectively.
    (c)~Variance of the shear rate increments as a function of time lag $\Delta t$ for the entire time series (black diamonds) together with the variances $\langle\delta \dot\gamma_{q}^2\rangle$ (blue circles) and $\langle\delta \dot\gamma_{p}^2\rangle$ (red squares) restricted to quiescent phases and to propagating events respectively. The thick gray line is the reconstruction of $\langle\delta \dot\gamma^2\rangle$ based on a linear combination $\alpha\langle\delta \dot\gamma_{q}^2\rangle+\beta\langle\delta \dot\gamma_{p}^2\rangle$ with $\alpha=0.82$ and $\beta=0.60$. The dotted and dashed lines have slopes 1 and 2 respectively.
     (d)~Local velocity $v(z,t)$ normalized by the rotor velocity $v_{\rm Rheo}(t)$ and averaged over isophases $t+z/c={\rm cst}$ along the direction ${\bf T}$ of band propagation with $c=11$~mm.s$^{-1}$. 
     (e)~Normalized radial velocity profiles $v(r)/v_{\rm Rheo}$ averaged at different stages of band propagation along the colored lines indicated in (b) (middle panel) and (d).}
	\label{fig:events_12pa_80pa}
\end{figure*}

A direct comparison between the shear rate response and local velocity data recorded just above DST onset ($\sigma=12$~Pa) is shown over long periods in Figs.~\ref{fig:SI3bis_spatiotemp_normed_12pa} and \ref{fig:SI2_velocity_12pa} in Appendix~\ref{s:spatio} and over the two recurring spatiotemporal patterns observed throughout regime I in Fig.~\ref{fig:events_12pa_80pa}, see also Supplemental Video~\ref{vid:12pa}. A first important observation is that in spite of strong slippage at both walls, the dispersion always remains {\it homogeneously} sheared in the bulk i.e. we do not observe any solidlike, shear-jammed state. A more thorough analysis of flow profiles, detailed in Appendix~\ref{s:velo} and displayed in Fig.~\ref{fig:SI2_velocity_12pa}, shows that the effective shear rate in the bulk suspension $\dot\gamma_B(t)$ correlates very well with $\dot\gamma(t)$. Therefore the global shear rate most likely reflects the local dynamics at least close to DST onset.

A second, even more striking piece of information that can be extracted from velocity maps in regime I is the presence of  {\it propagating events} along the vorticity direction that alternate with {\it quiescent} phases. Indeed, from the one-hour experiment at $\sigma=12$~Pa, we could extract $\sim$20 sequences of duration $\sim$10~s where the fluctuations of $\dot\gamma(t)$ and $v(r,z,t)$ remain essentially flat (Fig.~\ref{fig:events_12pa_80pa}a) and a similar number of sequences where $v(r,z,t)$ provides clear evidence for the propagation of a band along $z$ that is delineated in time by peaks in both the global and effective shear rates (Fig.~\ref{fig:events_12pa_80pa}b). As seen in Fig.~\ref{fig:SI3bis_spatiotemp_normed_12pa}, such global peaks may be present in both $\dot\gamma(t)$ and $v(r,z,t)$ with no sign of propagating events in between. This is of course because ultrasound imaging only captures a vertical slice of the sample and can miss traveling bands that do not span the whole azimuthal direction. Therefore, we shall now restrict the statistical analysis proposed above in Section~\ref{s:analysis} to sequences where these two different patterns can be unambiguously identified. 

Interestingly, during quiescent phases, the variance $\langle \delta\dot\gamma_q^2\rangle$ of the shear rate increments exhibits a purely diffusive scaling with time lag $\Delta t$ (blue symbols in Fig.~\ref{fig:events_12pa_80pa}c), which implies that the suspension applies random kicks on the surface of the rotating cylinder. On the other hand, the variance $\langle \delta\dot\gamma_p^2\rangle$ corresponding to sequences that include propagating events displays a short-time ballistic scaling up to $\Delta t\simeq 0.5$~s (red symbols in Fig.~\ref{fig:events_12pa_80pa}c). This characteristic timescale corresponds to the duration of the acceleration phases during global peaks in $\dot\gamma$ that precede and follow propagating events. Ultrasound images further show that propagating events travel along the vorticity direction with a speed $c\simeq 10$~mm\,s$^{-1}$. They extend over roughly 10~mm in the $z$-direction while they span the whole gap in the $r$-direction, see also Supplemental Video~\ref{vid:12pa}. Such traveling bands are associated with a peculiar temporal signature in the local velocity as shown in Fig.~\ref{fig:events_12pa_80pa}d. Within the band, the suspension flows at a velocity that is about 20\% smaller than average. The band is itself preceded by a front where the suspension moves about 10\% faster than average. Remarkably, the velocity profiles always show homogeneous shear across the gap, see Fig.~\ref{fig:events_12pa_80pa}e. The amount of wall slip is observed to transiently increase (decrease resp.) at the rotating (fixed resp.) cylinder while the effective shear rate in the bulk remains roughly constant.

Finally, coming back to the global shear rate dynamics, we use the two patterns associated with quiescent phases and propagating events as a projection basis to reconstruct the statistics of the whole $\dot\gamma(t)$ time series. As shown in Fig.~\ref{fig:events_12pa_80pa}c, a linear combination of $\langle \delta\dot\gamma_{q}^2\rangle$ and $\langle \delta\dot\gamma_{p}^2\rangle$ provides an excellent fit of the global variance $\langle\delta \dot\gamma^2\rangle$. Thus, singling out those two independent patterns based on local velocity maps allows us to recover the full dynamics from only about 10\% of the time series. 

\begin{figure}
	\centering
    \includegraphics[scale = 1]{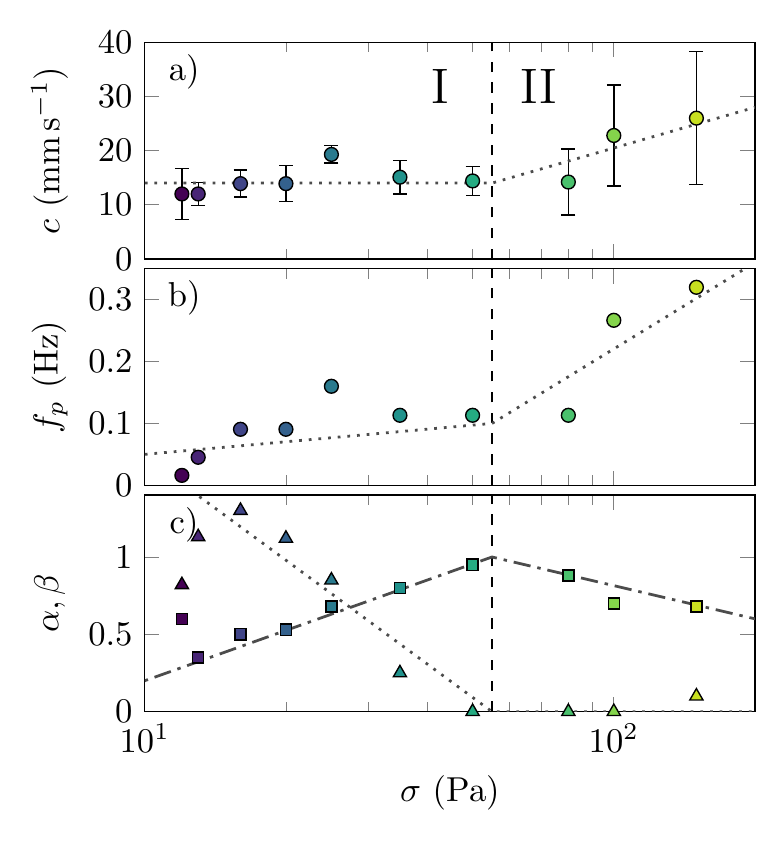}
    \caption{Characterization of the dynamics as a function of the applied stress $\sigma$. (a)~Speed $c$ of propagating events. Error bars correspond to the standard deviation taken over all detected events. (b)~Occurrence frequency $f_p$ of propagating events. (c)~Coefficients $\alpha$ (triangles) and $\beta$ (squares) of the best fit of $\langle \delta\dot\gamma^2\rangle$ by a linear combination $\alpha\langle \delta \dot\gamma_{q}^2 \rangle+\beta\langle \delta \dot\gamma_{p}^2\rangle$ of the variances of the two patterns shown in Fig.~\ref{fig:events_12pa_80pa}c. In the fitting procedure, the variance of the increments is normalized by the variance of the shear rate so that the coefficients can be compared from one shear stress to the other.}
	\label{fig:event_stats}
\end{figure}

The same analysis was repeated for stresses spanning both regimes I and II. Figure~\ref{fig:event_stats}a shows that the propagation speed of traveling bands stays roughly constant throughout regime I but shows large variability around increasing mean values in regime II. Meanwhile, the occurrence frequency of propagating events weakly increases in regime I before jumping to large values that are comparable to the cutoff frequencies observed in the PSDs in regime II, see Fig.~\ref{fig:event_stats}b. Strikingly, linear combinations of the variances $\langle \delta\dot\gamma_q^2\rangle$ and $\langle \delta\dot\gamma_p^2\rangle$ extracted at $\sigma=12$~Pa remain very efficient to reconstruct $\langle \delta\dot\gamma^2\rangle$ for all imposed stresses, see Fig.~\ref{fig:fits} in Appendix~\ref{s:stat}. This suggests that propagating bands interact very weakly with each other. As seen in Fig.~\ref{fig:event_stats}c, quiescent (i.e. diffusive) patterns dominate the dynamics in regime I whereas propagating (i.e. ballistic) events proliferate and eventually solely account for the dynamics in regime II. The latter proliferation of propagating events is directly illustrated through ultrasonic imaging at $\sigma=80$~Pa in Fig.~\ref{fig:SI3bis_spatiotemp_normed_80pa} in Appendix~\ref{s:spatio}, see also Supplemental Video~\ref{vid:80pa}. This proliferation also goes along with a decorrelation of the global shear rate relative to local velocities, see Fig.~\ref{fig:SI2_velocity_80pa}. The fact that the statistics of $\dot\gamma$ become Gaussian in regime II also supports a picture where propagating bands are independent random events.

\section{Discussion and interpretation in terms of unstable vorticity bands}

Our results demonstate that the unsteady dynamics observed in the DST of a dense cornstarch suspension originates from localized, intermittent propagating events that proliferate as the shear stress is increased. While vertical (or S-shaped) sections of the flow curve are usually attributed to steady-state banding \cite{Olmsted:2008}, we do not observe any shear localization involving solidlike arrested regions, unlike in experiments of Refs.~\cite{fall2015,hermes2016,rathee2017}. This is most probably because these previous works explored the shear-jammed region of the flow phase diagram \cite{wyart2014} either due to higher volume fractions \cite{rathee2017,hermes2016} or due to imposing the shear rate in a wide-gap Taylor-Couette geometry which led to particle migration and flow separation \cite{fall2015}. Here, we observe that the flow always remains homogeneously sheared in the gradient direction. Moreover we do not observe any signature of strong local variations of the volume fraction. The curvature of the velocity profiles seen in Fig.~\ref{fig:events_12pa_80pa}e, indicative of apparent shear-thinning, could be attributed to a mild migration towards the fixed outer cylinder, consistently with recent concentration measurements through X-ray imaging \cite{chatte2018}. Still, migration-induced local jamming cannot account for our experimental findings.

Based on the above observations, we may thus interpret the present results in terms of a homogeneous DST system that essentially undergoes an unsteady yet uniform shear rate. In such a case, the only physical parameter that can be invoked to explain instability is the local stress through, e.g., the presence of vorticity bands or stress-bearing structures. Of course, our experimental setup does not provide access to local stresses in the bulk suspension but we shall argue below that the observed propagating events are fully consistent with traveling vorticity bands that were predicted and observed numerically very recently \cite{chacko2018}. 

Theoretical arguments rule out the possibility of steady-state vorticity bands for hard non-Brownian particles \cite{hermes2016}. Indeed, were vorticity bands stable, both the particle pressure and the solvent pressure would need to balance separately in order to stabilize the band interface. This would lead to identical shear stresses and thus to the absence of bands. Rather, DST flows were predicted to be always unsteady and a two-dimensional extension of the Wyart and Cates model was shown to yield oscillatory and chaotic-like dynamics in frictional granular particles \cite{grob2016} calling for three-dimensional approaches. Such three-dimensional simulations were performed recently by Chacko {\it et al.} \cite{chacko2018} who also devised a one-dimensional instability model for vorticity bands.

Both the model and the simulations in Ref.~\cite{chacko2018} predict shear-thickened bands bearing a shear stress larger than average that travel along the vorticity direction at a given speed $c$. Simulations show that the band moves by about 800 particle radii per strain unit. With our cornstarch grains of mean diameter 15~$\mu$m (Appendix~\ref{s:sample}) and a typical strain rate of 2~s$^{-1}$, this corresponds to $c\simeq 12$~mm\,s$^{-1}$ in striking quantitative agreement with Fig.~\ref{fig:event_stats}a. Although simulations still remain to be made fully comparable to experiments, in particular concerning boundary conditions, this agreement strongly suggests that our propagating events correspond to the traveling bands of Ref.~\cite{chacko2018}.

\begin{figure}
	\centering
    \includegraphics[scale=1]{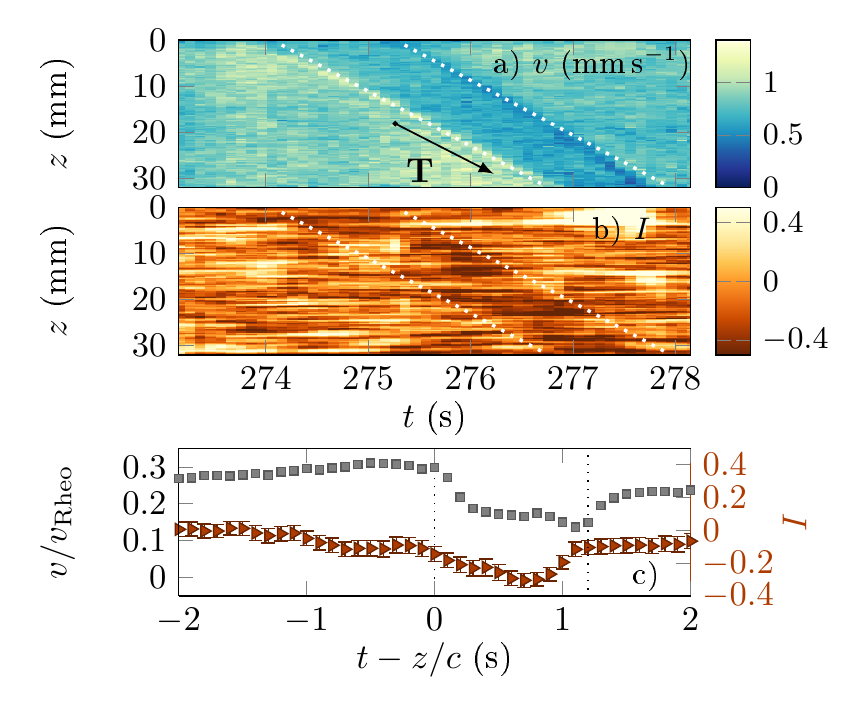}
    \caption{Signature of a propagating event in regime II for $\sigma = 80$~Pa. Spatiotemporal maps of (a)~the local velocity $v(r,z,t)$  and (b)~the ultrasonic intensity $I(r,z,t)$ averaged in the $r$-direction. (c)~Normalized velocity $v/v_{\rm Rheo}$ (grey) and intensity $I$ (orange) averaged over isophases $t-z/c={\rm cst}$ along the direction ${\bf T}$ of band propagation with $c=12.2$~mm.s$^{-1}$.}
	\label{fig:Event_80Pa}
\end{figure}

Another prediction by Chacko {\it et al.} \cite{chacko2018} is that band propagation goes along with small local variations of the vertical velocity $v_z$ and of the volume fraction $\phi$. Although we cannot demonstrate the presence of non-zero vertical velocities due to the low resolution of our ultrasonic imaging setup in the $z$-direction, we may get some hints for local variations of the density by focusing on the local ultrasonic intensity $I$. Indeed, using the same setup on various non-Brownian suspensions in the dilute and semidilute regimes, we have shown that $I$ directly relates to $\phi$, yet in a complex manner for suspensions above a few volume percent \cite{saint2017b}. While intensity maps remain featureless in regime I at $\sigma=12$~Pa (see Fig.~\ref{fig:SI3bis_spatiotemp_normed_12pa}d), a convincing correlation between the tangential velocity and the local intensity is reported in Fig.~\ref{fig:Event_80Pa} for $\sigma=80$~Pa in regime II. There, propagating events are more pronounced and we observe that the intensity varies in the band concomitantly with the local velocity, see also Fig.~\ref{fig:SI3bis_spatiotemp_normed_80pa}d for more events. Therefore, our experiments are consistent with the simulations that display an increased volume fraction at the front of the vorticity bands followed by a small depletion within the band. More quantitative local measurements of the volume fraction, e.g. through time-resolved X-ray imaging, are in line to strengthen the present qualitative observations.

To push the discussion further, we note that the interpretation of propagating events in terms of unstable vorticity bands is also supported by the behaviors of the local velocity and wall slip shown in Fig.~\ref{fig:events_12pa_80pa} during an event. If the volume fraction is slightly larger at the outer cylinder, as indicated by the curvature of the velocity profiles, then one expects the network contacts to exert a larger stress there in the case of sliding friction at the walls. Equivalently, in the case of lubricated friction, one expects the lubricating layer to be compressed and thinner at the fixed wall. This would result in a lower slip velocity at the stator during the propagation of a band bearing a higher local stress. Thus we may speculate that slippage of the suspension indirectly probes the local stress. We also note that the events unveiled in the present study clearly differ from large-amplitude stick-slip oscillations that may couple to the rheometer inertia as reported in other works focusing on shear-jammed DST suspensions \cite{Larsen:2014,Bossis2017}. However, wall slip during DST of dense suspensions definitely requires more attention in future work especially to elucidate its microscopic origin and its possible interplay with the unstable bulk dynamics. 

Finally, our results raise a number of interesting open issues. First, we observed that propagating events go along with sudden acceleration and deceleration of the rotating cylinder. These global peaks are likely to correspond to fast elastic waves generated during the nucleation of the bands or from their interactions with the bottom wall and/or the top surface. Conversely, the fact that these peaks are not systematically associated with a propagating event detected at the probe location suggests that the bands do not span the whole circumference of the cell. This points to the need to simulate extended yet bounded three-dimensional systems in order to investigate how the bands precisely nucleate and what sets their spatial extent. 3D simulations of dense suspensions were recently performed based on a fluid dynamics model for dilatant fluids  where the inertia of the fluid plays a central role \cite{nakanishi2012,Nagahiro:2016}. This approach accounts for the large and fast ``shear thickening oscillation'' observed in the wide-gap Taylor-Couette flow of potato-starch suspensions \cite{Nagahiro:2013} and predicts localized shear-thickened bands that bear negative pressures but that do not seem to organize nor propagate along the vorticity direction \cite{Nagahiro:2016}. 
The case of a small-gap Taylor-Couette flow with much smaller inertia and more homogeneous shear rate and volume fraction conditions still remains to be explored numerically as well as experimentally through local wall or particle pressure measurements \cite{deboeuf2009,boyer2011,garland2013}.

Second, we successfully rationalized the complex DST dynamics of cornstarch based on two different spatiotemporal patterns extracted from a statistical analysis inspired by turbulence studies. This approach was only possible thanks to long experiments and to local insights into the flow. We do not however report any clear oscillatory dynamical regime such as the periodic and chaotic-like states in the experiments of Ref.~\cite{hermes2016} or the ``locally oscillating bands'' in the model and simulations of Ref.~\cite{chacko2018}. This could be due to differences in the location of the system within the shear-thickening phase diagram. Flow curves measured at different cornstarch weight fractions are shown in Fig.~\ref{fig:concentration} in Appendix~\ref{s:concentration} in order to facilitate the comparison with other results. We also note that the systems investigated in Refs.~\cite{hermes2016,chacko2018} are more confined than in the present work. The possibility of truly chaotic dynamical regions in the DST phase diagram should thus be ascertained together with their dependence on geometry. To stimulate such a future study, we show in Fig.~\ref{fig:geometry} in Appendix~\ref{s:geo} that fluctuations similar to those reported above are also observed in two other geometries. Yet, it is highly probable that the nature of the instability and its spatiotemporal signature depend on a number of physical degrees of freedom including shearing geometry, sample size and boundary conditions. 

Third, our statistical approach has revealed a specific range of timescales where the dynamics is diffusive-like in regime I. Such a separation of timescales is reminiscent of multiscale processes, e.g. hydrodynamic turbulence \cite{castaing1990} or imbibition processes \cite{clotet2014}. In the present case of shear-thickening flows, however, it is still unclear whether some kind of self-similarity is involved. In particular, the traveling bands unveiled here could represent an analogue of large-scale coherent structures in turbulent flows but the available statistics does not allow us to tell whether they present some hierarchy of scales. Moreover, although the short timescale was related to the duration of global peaks in the mechanical response, the nucleation mechanism of the bands and what exactly sets this timescale are unknown. Along the same lines, the timescale above which the statistics become fully Gaussian is probably linked to the lifetime of the bands, to their propagation speed and/or to their occurrence frequency but the processes that select such characteristics are still to be discovered.

\section{Conclusion}

By carefully investigating a dense cornstarch suspension with a combination of rheometry and ultrasound imaging over long time scales, we have pinned the origin of unsteady dynamics in DST to the existence of transient localized bands that travel along the vorticity direction. The global dynamics can be decomposed into ballistic phases that correspond to such traveling bands and diffusive phases that correspond to a background of random kicks applied to the moving tool. As the shear stress is increased in the DST regimes, propagating bands progressively dominate over diffusive phases, leading to Gaussian fluctuations. These key observations urge to introduce stress spatiotemporal correlations in models and to simulate dense three-dimensional systems together with the shearing boundaries. In particular, the striking similarities between our experiments and recent model and simulations strongly support an explanation of shear-thickening dynamics based on localized vorticity bands bearing high shear stresses. Further experimental work on DST suspensions should now focus on the measurements of bulk stresses to directly confirm this picture as well as more microscopic observations including sliding of the material at the walls.

Our results concretize recent scientific effort to unravel shear-thickening in dense suspensions and to understand its control through the addition of plasticizers~\cite{Lombois-Burger2008,Bossis2017,Abdesselam2017}, nanofibers~\cite{ghosh2017} or large particles~\cite{madraki2017} and the application of magnetic fields~\cite{Bossis2016} or transverse shear~\cite{lin2016}. Our work may also shed new light on undesirable effects in industrial processes such as stress surges~\cite{lootens2003,Perrot2006}, granulation and slurry fracture~\cite{Beazley1965,Cates:2005} and help to further develop promising applications such as fabrics impregnated with shear-thickening fluids for improved resistance to impact~\cite{Lee2003,Sun:2013,Cwalina:2016,Gurgen2017}.

\section*{Acknowledgements} BSM thanks Solvay for funding. This research was supported in part by the National Science Foundation under Grant No. NSF PHY-1748958 through the KITP program on the Physics of Dense Suspensions. The authors are grateful to Laurent Chevillard for fruitful advice on signal analysis. We deeply thank Romain Mari and Raul Chacko for enlightening discussions on models and simulations of shear-thickening systems, as well as Mike Cates and Suzanne Fielding for sharing their latest results with us. We also wish to acknowledge insightful discussions with D.~Blair, M.~Bourgoin, A.~Colin, J.~Comtet, M.~Denn, T.~Divoux, E.~Han, H.~Jaeger, G.~McKinley, P.~Olmsted, G.~Ovarlez, A.~Sood and M.~Wyart at various stages of this work.

\bibliography{biblio}

\clearpage
\appendix

\begin{widetext}

\section{Supplemental information}


\subsection{Cornstarch suspension} 
\label{s:sample}
Dense cornstarch suspensions are obtained by dispersing 41\% wt. of cornstarch (Sigma-Aldrich, CAS 9005-25-8, S4126-2KG) at room temperature in a density-matched solvent composed of 46\% wt. water and 54\% wt. cesium chloride (CsCl) \cite{Han2016}. Figures~\ref{fig:Maizena41_recap_20170922_20} and \ref{fig:cornstarch} respectively show density measurements of water--cesium chloride mixtures and the cornstarch size distribution inferred from optical microscopy. Cornstarch particles have a mean diameter $\langle a \rangle =15~\mu$m and the standard deviation of the diameter distribution is $7~\mu$m. Cornstarch is progressively added to the solvent and lumps of starch are broken using a spatula. The suspension is then centrifuged at 200~$g$ for two minutes before being poured into the Taylor-Couette cell. Following Ref.~\cite{Han2016}, the weight fraction used in the present work corresponds to a volume concentration of about 47\%. To remove trapped air bubbles that may affect the rheology of the cornstarch suspension, preshear is applied for at least $5$~h under a low shear stress $\sigma=5$~Pa. As seen in Fig.~\ref{fig:SI1_bubbles}, the suspension initially scatters ultrasound very strongly due to air bubbles. Shearing the suspension for several hours in the shear-thinning regime below shear-thickening allows us to progressively remove air bubbles, as confirmed by the slow decrease in the backscattered ultrasound intensity. The speckle signal obtained after $\sim 5$~h is homogeneous and corresponds to backscattering by the cornstarch grains with a negligible contribution from air bubbles (if any).

\begin{figure}[h]
	\centering
    \includegraphics[scale = 0.85]{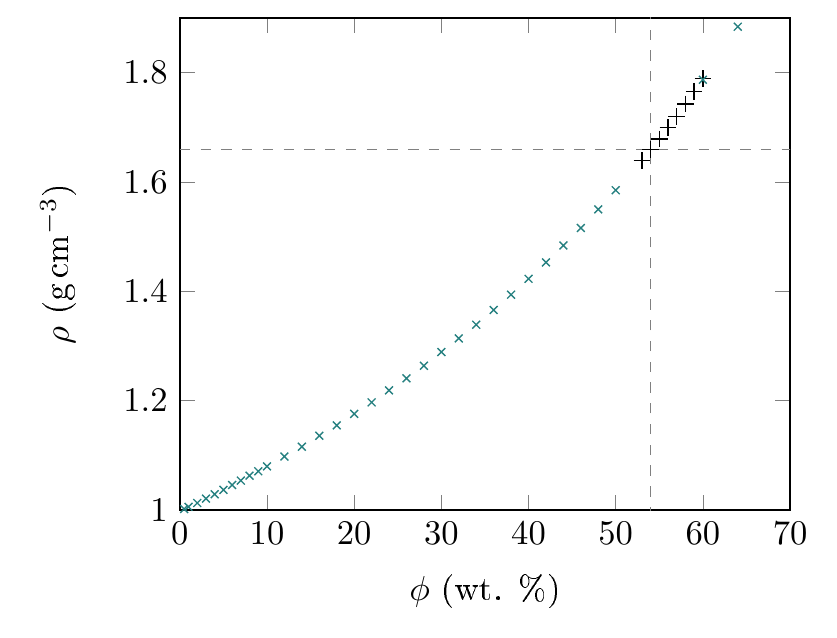}
    \caption{Density of water--cesium chloride mixtures as a function of the cesium chloride mass fraction $\phi$. The data below $\phi=50$\% wt. was retrieved from Ref.~\cite{Handbook}. Measurements above  $\phi=50$\% wt. were performed at a temperature of 22$^\circ$C using an acoustic resonance densimeter (Anton Paar DMA 35). Dotted lines show that matching the cornstarch density of 1.63 requires $\phi=54$\% wt.}
	\label{fig:Maizena41_recap_20170922_20}
\end{figure}

\begin{figure*}[h]
	\centering
    \includegraphics[scale = 0.85]{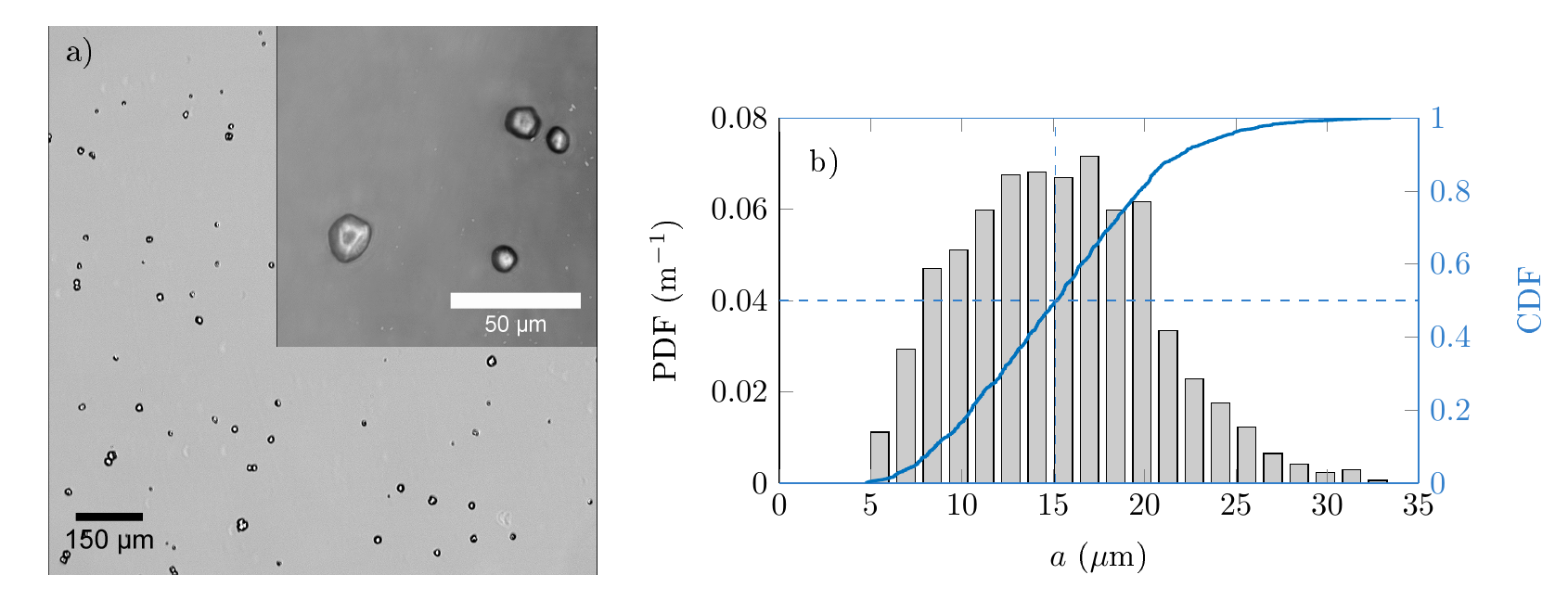}
    \caption{Particle size distribution of cornstarch. (a) Bright field microscopy image of a dilute suspension of cornstarch. (b)~Particle diameter distribution obtained from microscopy images (in gray) and cumulative distribution function (in blue). Dotted lines indicate the mean particle diameter $\langle a \rangle =15~\mu$m.}
	\label{fig:cornstarch}
\end{figure*}

\clearpage

\begin{figure*}[h]
	\centering
    \includegraphics[scale = 0.9]{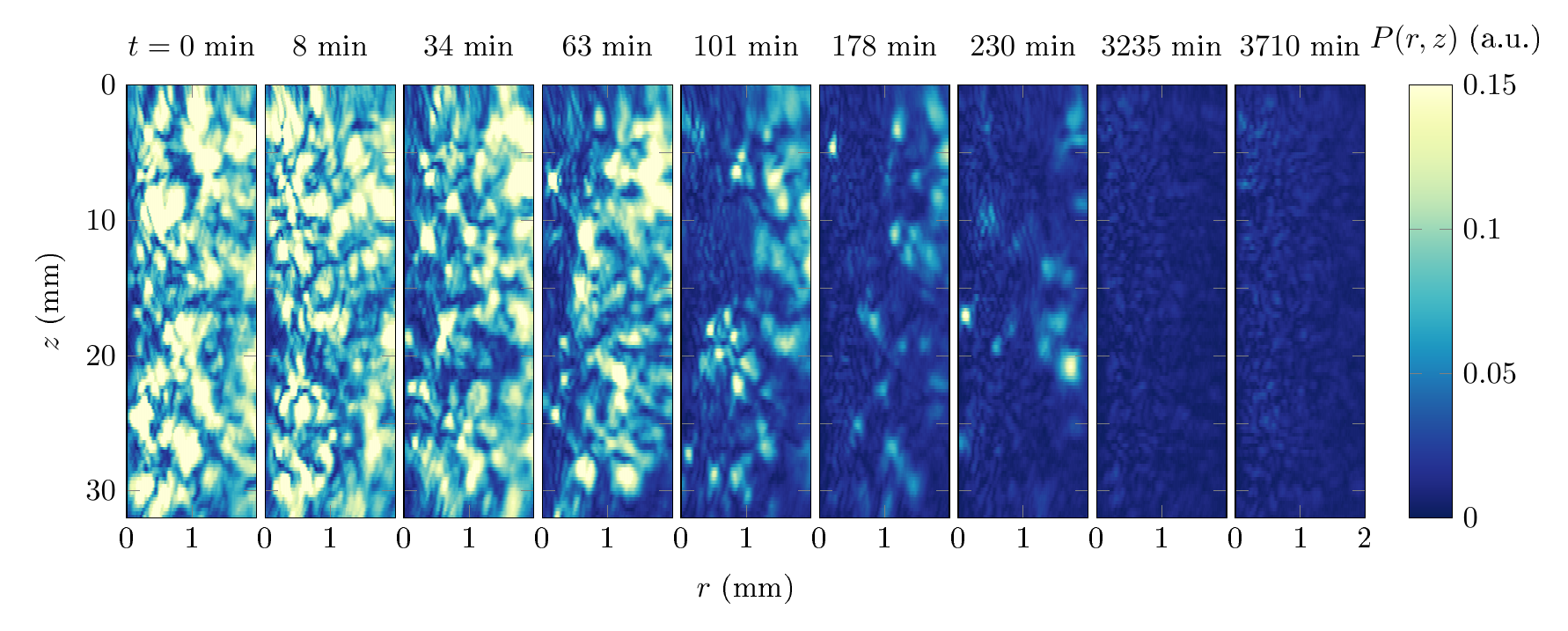}
    \caption{Removing air bubbles trapped in the cornstarch dispersion. Maps of the amplitude $P(r,z)$ of the ultrasonic pressure signal backscattered by the cornstarch suspension at various times during preshear under a shear stress $\sigma=5$~Pa. The bright spots in the images correspond to air bubbles that are initially trapped but progressively rise through the sheared suspension until no bubble is seen across the ultrasound region of interest after about 5~h.}
	\label{fig:SI1_bubbles}
\end{figure*}

\subsection{Experimental setup}
\label{s:rheo}
Figure~\ref{fig:setup_pics} shows a picture of our experimental setup. Rheological measurements are carried out in a concentric-cylinder (or Taylor-Couette) geometry driven by a stress-controlled rheometer (TA Instruments ARG2). The fixed outer cylinder (stator) has an inner radius $R_o=25$~mm and the rotating inner cylinder (rotor) an outer radius $R_i = 23$~mm, leaving a gap $e=R_o-R_i=2$~mm between the two cylinders. In such a small-gap Taylor-Couette geometry, the curvature induces a relative decrease of the shear stress by $\delta\sigma/\sigma=(R_o/R_i)^2-1=0.18$ from the inner cylinder to the outer cylinder. Let us emphasize that the stress heterogeneity inherent to our cell is twice smaller than in the Taylor-Couette geometry used by Hermes {\it et al.} \cite{hermes2016}, for which $\delta\sigma/\sigma=0.36$, and ten times smaller than in the wide-gap geometry of Fall {\it et al.} \cite{fall2015} where $\delta\sigma/\sigma=1.8$, i.e. where the stress is almost three times smaller at the stator than at the rotor. Shear is even more heterogeneous in a parallel-plate geometry where the shear rate goes from zero on the axis of rotation to its maximum value at the periphery \cite{hermes2016}. Therefore, our small-gap Taylor-Couette geometry ensures a good homogeneity of the shear field, which helps a lot in mitigating particle migration due to shear gradients.

The gap width $e\simeq 130\langle a \rangle$ is still large compared to the mean particle size such that no significant effect of confinement is expected. Moreover, our cell has a large aspect ratio thanks to its height $H=63$~mm $\simeq 30e\simeq 4,\!000 \langle a \rangle$. This large aspect ratio corresponds to a small ratio of the free surface of the suspension to the rotating cylinder surface over which the total torque measured by the rheometer is integrated, $S_{\rm free}/S_{\rm tool}\simeq e/H\simeq 0.03$. For typical cone-and-plate or parallel plate devices with radius 20~mm and gap width 1~mm at the periphery, the ratio is $S_{\rm free}/S_{\rm tool}\simeq 0.1$, which makes those geometries more sensitive to instabilities of the free surface.

Both cylinders are made of smooth Delrin (polyoxymethylene). This material was chosen because it leads to limited slippage compared to, e.g., smooth PMMA surfaces. However, as discussed in the main text, wall slip has a deep connection to the DST bulk dynamics and certainly deserves more attention in itself. In particular, rather than a mere artifact that has to be eliminated, it should be treated as a complex yet interesting physical phenomenon that may carry key information on the system under study. Moreover the inertia due to the geometry and to the rheometer can be neglected in all experiments. Indeed our setup has a total moment of inertia $I_{\rm tot}=54~\mu$N\,ms$^2$ and the maximum strain acceleration was measured to be $\ddot{\gamma}_{\rm max}\simeq 8$~s$^{-2}$ so that the corresponding stress is at most $\sigma_{\rm i,max}=eI\ddot{\gamma}_{\rm max}/(2\pi H R_i^3)\simeq 0.18$~Pa, always below 2\% of the imposed shear stress. The fluid has a moment of inertia $I_{\rm fluid}=\pi\rho H(R_o^4-R_i^4)/2\simeq 18~\mu$N\,ms$^2$ and brings an even smaller contribution to inertial stresses. The Reynolds number, ${\rm Re} = \rho e^2 \dot\gamma / \eta$, never exceeds $10^{-2}$. We may therefore neglect inertia in our stress measurements. 

Finally, the stator is closed by a lid as sketched in Fig.~\ref{fig:setup_sketch}. A groove machined in the upper surface of the rotor is filled with water to act as a solvent trap and minimize evaporation. This allows us to perform reproducible experiments on a time span of more than 20~hours with the same loading of the cell. We used the ARG2 Auxiliary Sample utility to retrieve the applied stress and the shear rate response as a function of time with a sampling frequency of 500~Hz.

\begin{figure*}[htb]
	\centering
    \includegraphics[scale = .32]{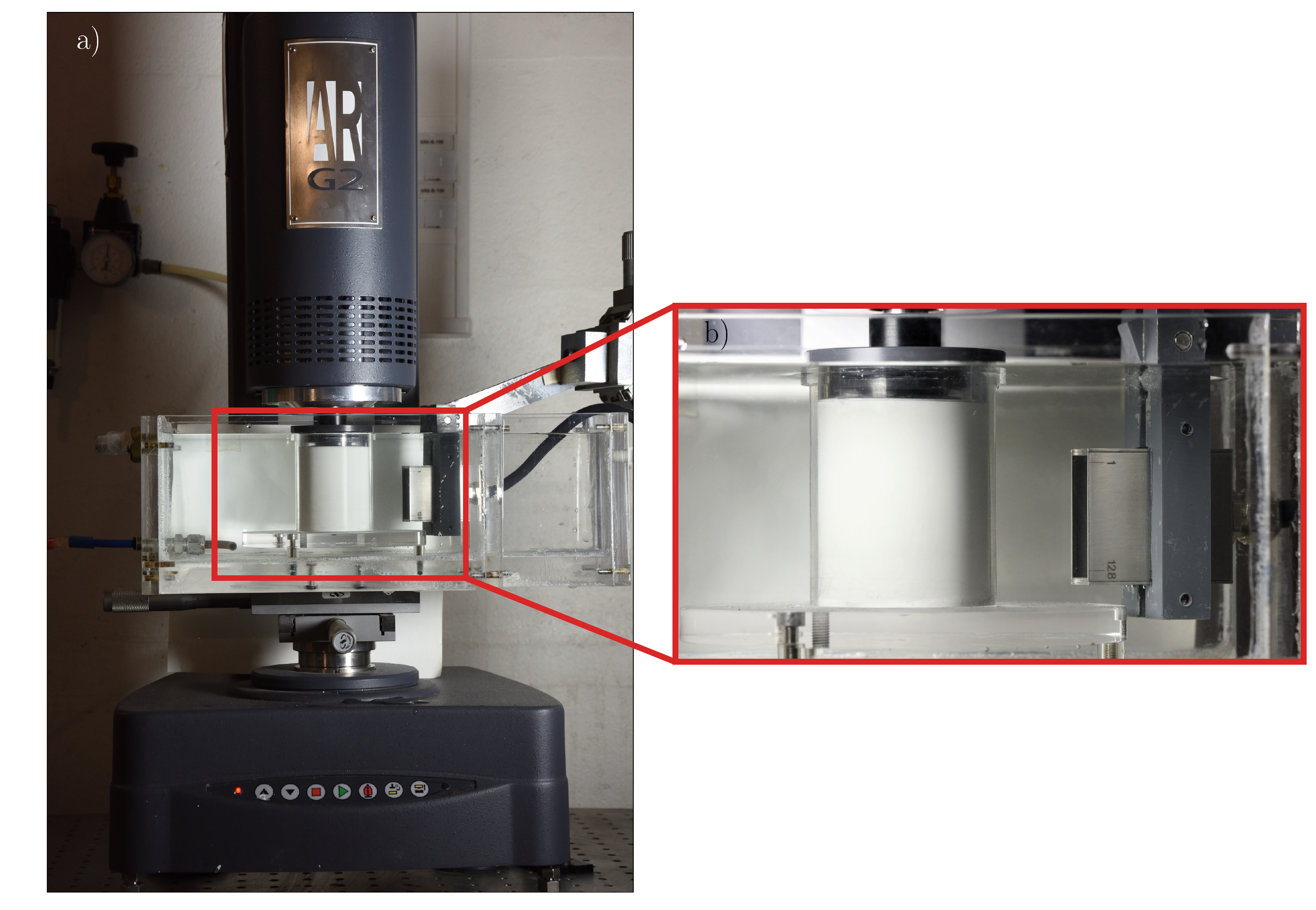}
    \caption{Picture of the experimental arrangement. (a)~Global view of the setup showing the rheometer with the large water tank surrounding the Taylor-Couette cell and used both for thermostatation and for acoustic impedance matching between the cell and the ultrasonic probe. (b)~Enlargement showing the ultrasonic probe (on the right) facing the Taylor-Couette geometry filled with the cornstarch suspension and closed by a lid (in gray) to minimize evaporation.}
	\label{fig:setup_pics}
\end{figure*}

\begin{figure*}[h!]
	\centering
    \includegraphics[scale = .65]{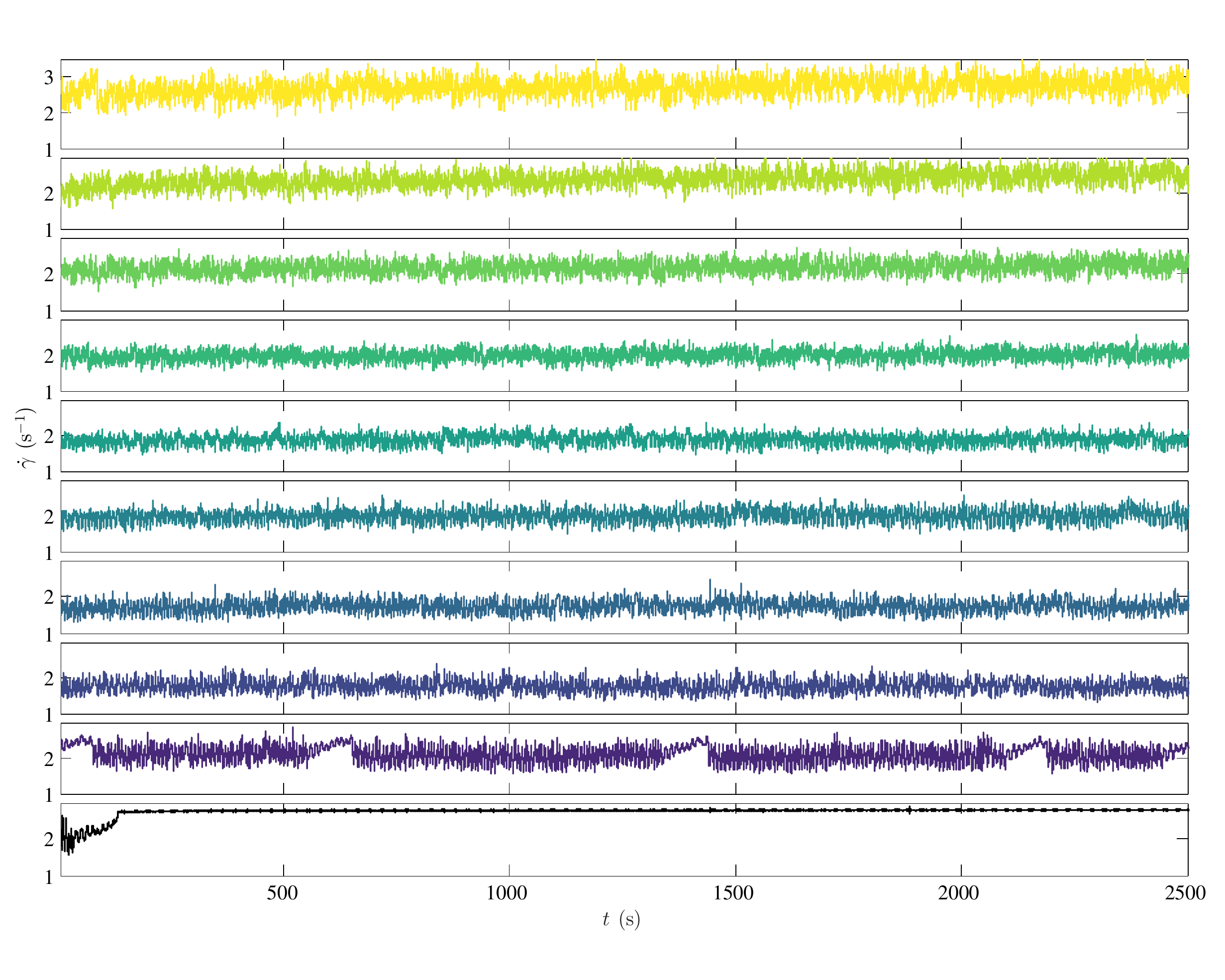}
    \caption{Time series of the shear rate $\dot\gamma(t)$ recorded under various constant shear stresses imposed at time $t=0$: $\sigma=11$~Pa up to 150~Pa from bottom to top (see color code in Fig.~\ref{fig:fc_tseries_spectra}).}
	\label{fig:Maizena41_recap_20170922_28}
\end{figure*}

\clearpage

\subsection{Signal analysis}
\label{s:stat}

Long experiments under a constant stress in the DST regime yield statistically stationary shear rate responses $\dot\gamma(t)$. Figure~\ref{fig:Maizena41_recap_20170922_28} shows a selection of such responses on more than 2,000~s. To characterize the temporal fluctuations of the shear rate, we use the following tools borrowed from studies of turbulent signals \cite{castaing1990,chevillard2012}.

\begin{itemize}

\item The power spectrum density (PSD) of $\dot\gamma(t)$ is defined as 
$ {\rm PSD}\{\dot\gamma\}=\mathcal{F} \{ \dot\gamma \} \mathcal{F}^* \{ \dot\gamma \}$, where $\mathcal{F} \{\cdot \}$ is the Fourier transform operator.

\item The probability distribution function (PDF) of the increments $\delta\dot\gamma (t,\Delta t) = \dot\gamma(t + \Delta t) - \dot\gamma(t)$ is the probability $
{\mathcal{P}}(\delta\dot\gamma,\Delta t)$ to find an event of amplitude $\delta \dot \gamma$ in the increment time series for a given $\Delta t$. By definition, for any given value of $\Delta t$, the mean of $\delta\dot\gamma (t,\Delta t)$ is zero.

\item The second moment of the PDF is the variance $\langle \delta\dot\gamma^2\rangle(\Delta t)=\langle\delta\dot\gamma (t,\Delta t)^2\rangle_t$, where the brackets $\langle\cdots\rangle_t$ denote the average over time $t$. 
As explained in the main text, whatever the applied stress in DST, the variance $\langle \delta\dot\gamma^2\rangle$ can be fitted by linear combinations of the variances of two elementary processes  $\langle \delta\dot\gamma_q^2\rangle$ and $\langle \delta\dot\gamma_p^2\rangle$ extracted just above DST onset. Such fits are shown in Fig.~\ref{fig:fits}.

\item Related to the third moment of the PDF, we introduce the skewness defined as \begin{equation}
{\mathcal{S}}(\Delta t) = \frac{ \langle \delta\dot\gamma(t,\Delta t)^3\rangle_t}{\langle\delta\dot\gamma^2\rangle^{3/2}}\,.
\end{equation}
${\mathcal{S}}=0$ indicates that the PDF is symmetric. ${\mathcal{S}}>0$ (${\mathcal{S}}<0$ resp.) means that the distribution is concentrated on the right (left resp.) side of $\langle \delta\dot\gamma \rangle$.

\item Finally, the fourth moment of the PDF is used to define the logarithm of the normalized kurtosis:
\begin{equation}
{\mathcal{K}}(\Delta t) = \ln\left( \frac{ \langle \delta\dot\gamma(t,\Delta t)^4\rangle}{3\langle\delta\dot\gamma^2\rangle^{2}}\right)\,.
\end{equation}
The normalization is such that ${\mathcal{K}}=0$ for a Gaussian distribution. ${\mathcal{K}}>0$ (${\mathcal{K}}<0$ resp.) indicates that outlier events are more (less resp.) probable than in a Gaussian distribution. 

\end{itemize}

\begin{figure*}[h!]
	\centering
	\includegraphics[scale = .78]{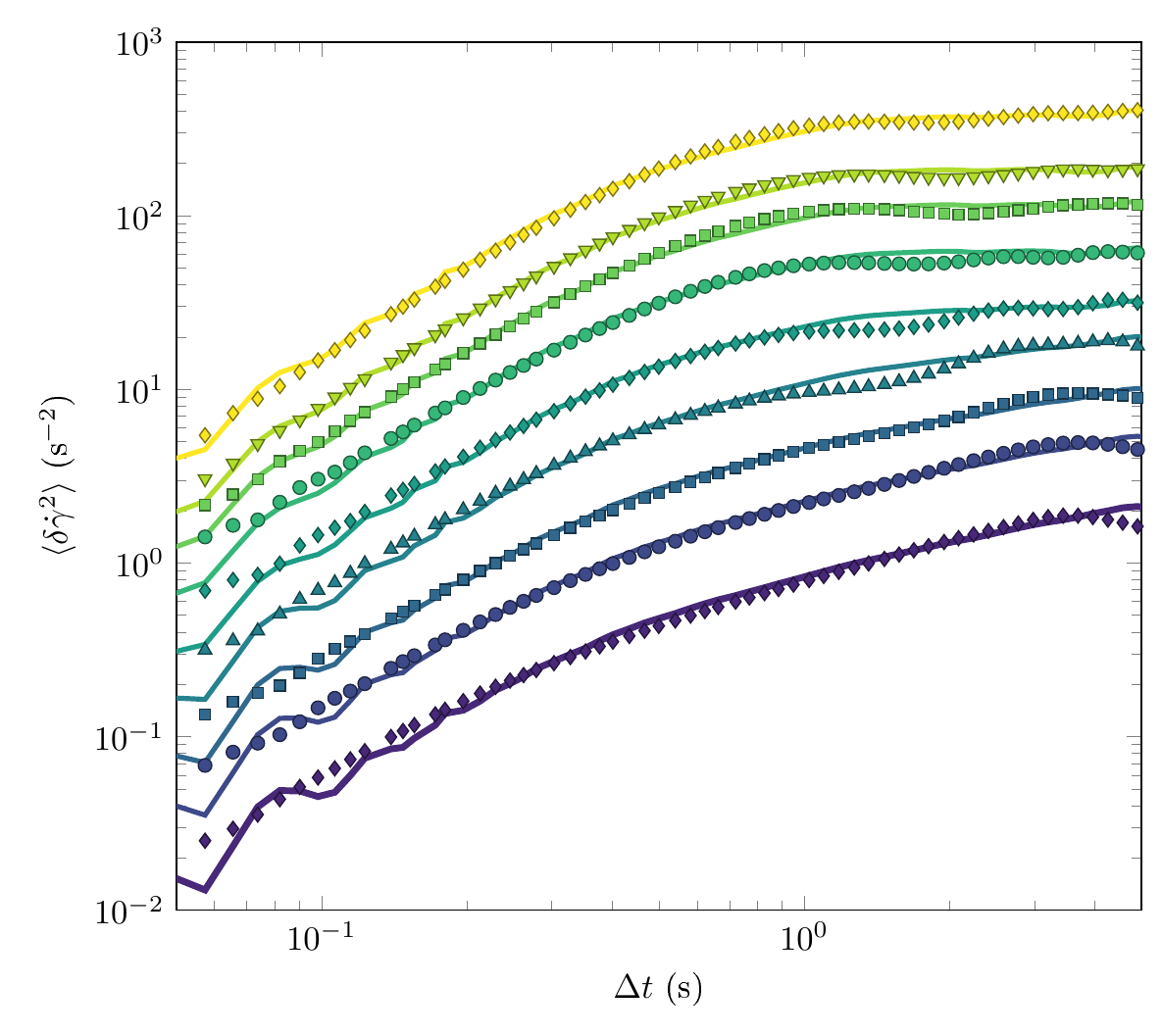}
    \caption{Variance $\langle \delta\dot\gamma^2\rangle(\Delta t)$ of the shear rate for various imposed stresses (colored symbols) together with the fits by linear combinations of $\langle \delta\dot\gamma_q^2\rangle$ and $\langle \delta\dot\gamma_p^2\rangle$ as detailed in the main text (colored lines). The coefficients $\alpha$ and $\beta$ used for the fits are shown in Fig.~\ref{fig:event_stats}c. For clarity, successive data sets were shifted vertically by 2 starting from $\sigma=13$~Pa.}
	\label{fig:fits}
\end{figure*}

\clearpage

\subsection{Ultrasound imaging}
\label{s:echo}
Rheological measurements are synchronized with ultrafast ultrasonic imaging~\cite{gallot2013}. Our echography technique relies on a custom-made high-frequency scanner driving an array of 128 piezoelectric transducers that sends short plane ultrasonic pulses with 15~MHz center frequency across the gap of the Taylor-Couette cell. As displayed in Fig.~\ref{fig:setup_pics}, the ultrasonic probe is immersed in a large water tank for acoustic impedance matching and is set vertically at about 25~mm from the stator. The water tank is connected to a circulation bath (Huber Ministat 125) that regulates the temperature to $T=25\pm 0.1~^{\circ}$C. The full specifications of this rheo-ultrasound setup can be found in Ref.~\cite{gallot2013}. During its propagation through the suspension, a plane ultrasonic pulse gets scattered by the cornstarch particles. The backscattered pressure signals constitute an ultrasonic ``speckle'' that is recorded by the transducer array, sampled at 160~MHz and further post-processed into images of the echogeneous structure.

Cross-correlating successive images yields the tangential velocity of the sample $v(r, z, t)$ at time $t$ and position $(r,z)$ in cylindrical coordinates, $r$ denoting the distance to the stator and $z$ the position along the vertical direction pointing downwards with the origin taken at transducer \#1 (see Fig.~\ref{fig:setup_pics}b). The ultrasonic images cover 32~mm in height, i.e. about half of the Taylor-Couette cell in the vorticity direction. Moreover, the characteristics of the ultrasonic beam corresponds to an azimuthal span of 300~$\mu$m. The spatial resolution along the $z$-direction is given by the spacing of 250~$\mu$m between two adjacent transducers and the resolution in the radial direction  $r$ is 75~$\mu$m. The frame rate for ultrasonic images is fixed to $50$~fps in all the experiments presented here.

Velocity measurements from ultrasound speckle images rely on the phase of the backscattered pressure signal. We can also measure the amplitude $P(r, z, t)$ of the backscattered signal through a Hilbert transform. As shown in Ref.~\cite{saint2017b} for a number of suspensions of non-Brownian particles with diameters ranging from 20 to 80~$\mu$m, $P(r, z, t)$ is a monotonic function of the local volume fraction $\phi(r,z,t)$. Here, we shall only use $P(r, z, t)$ as an indication of local variations of $\phi(r,z,t)$ since an absolute determination of the local volume fraction requires careful calibration and strong additional assumptions on scattering and attenuation by the suspension~\cite{saint2017b}.


\subsection{Spatiotemporal diagrams from $v(r,z,t)$ and $P(r,z,t)$}
\label{s:spatio}

In order to put emphasis on the dynamics of the local tangential velocity either along the vorticity direction $z$ or along the gradient direction $r$, spatiotemporal diagrams are computed by averaging $v(r,z,t)$ along the $r$-direction or the $z$-direction, which yields respectively the $v(z,t)$ and $v(r,t)$ maps shown in Figs.~\ref{fig:events_12pa_80pa}a,b, \ref{fig:Event_80Pa}a, \ref{fig:SI3bis_spatiotemp_normed_12pa}b,e and \ref{fig:SI3bis_spatiotemp_normed_80pa}b,e. Similarly, we present spatiotemporal diagrams $I(z,t)$ of the local ultrasonic intensity in Figs.~\ref{fig:Event_80Pa}b, \ref{fig:SI3bis_spatiotemp_normed_12pa}d and \ref{fig:SI3bis_spatiotemp_normed_80pa}d. There, in order to remove small yet systematic discrepancies in the intensity response of the different transducers, the intensity $I(z,t)$ is computed from $P(z,t)$, the $r$-average of $P(r,z,t)$, as $I(z,t)= (P(z,t) - \langle P(z,t)\rangle_t) / \sigma_{P(z)}$, where $\sigma_{P(z)}$ denotes the standard deviation of $P(z,t)$ taken over $t$.

Typical velocity and intensity spatiotemporal diagrams are displayed over the course of about 1~min in Fig.~\ref{fig:SI3bis_spatiotemp_normed_12pa} for $\sigma=12$~Pa (regime I) and in Fig.~\ref{fig:SI3bis_spatiotemp_normed_80pa} for $\sigma=80$~Pa (regime II). For the lower shear stress, we observe that the propagating event detected at $t\simeq 575$~s is preceded and followed by global peaks in the shear rate $\dot\gamma$, see Fig.~\ref{fig:SI3bis_spatiotemp_normed_12pa}a. These peaks also show as vertical lines along the $z$ and $r$ axes in the spatiotemporal maps of Fig.~\ref{fig:SI3bis_spatiotemp_normed_12pa}b,e. Based on the presence of various other peaks in $\dot\gamma$, we infer that during this specific time sequence, a few other propagating events take place yet outside the ultrasound region of interest. Global peaks disappear from the spatiotemporal representations when the local velocity is normalized by the rotor velocity $v_{\rm Rheo}(t)$, see Fig.~\ref{fig:SI3bis_spatiotemp_normed_12pa}c,f. This confirms that the propagating events are local while the sudden acceleration phases before and after the events are global.

For the larger shear stress, the local dynamics are decorrelated from the global ones. Yet, locally, propagating events are still delineated in time by maxima of the velocity that occur simultaneously for all $z$ and $r$. The fact that normalizing the local velocity by the rotor velocity does not significantly affect the spatiotemporal diagrams in this case (compare Fig.~\ref{fig:SI3bis_spatiotemp_normed_80pa}b,e with Fig.~\ref{fig:SI3bis_spatiotemp_normed_80pa}c,f) suggests that many different propagating events take place independently at the same time throughout the sample and average out in the global signals $v_{\rm Rheo}(t)$ and $\dot\gamma(t)$.

\begin{figure*}
	\centering
    \includegraphics[scale = 0.85]{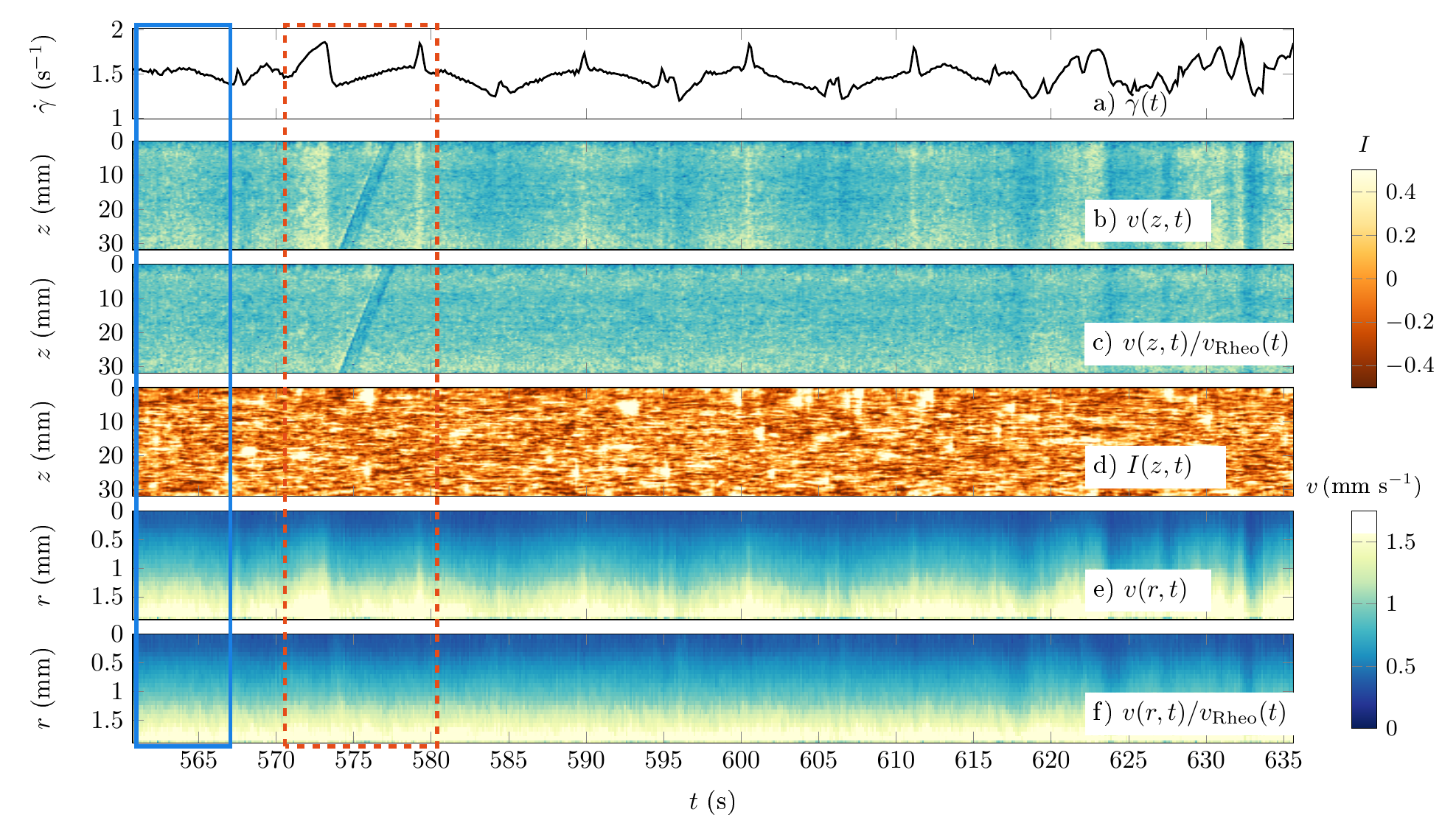}
    \caption{Comparison between global and local dynamics in regime I for $\sigma = 12$~Pa. (a)~Global shear rate response $\dot\gamma(t)$ recorded by the rheometer as a function of time. (b)--(f)~Spatiotemporal diagrams of the velocity $v(r,z,t)$ and of the ultrasonic intensity $I(r,z,t)$ as defined in Appendix~\ref{s:spatio}. The boxes highlight a diffusive, quiescent phase (blue) and the ballistic phase associated with a propagating event (red) discussed in the main text and enlarged in Fig.~\ref{fig:events_12pa_80pa}b. See also Supplemental Video~\ref{vid:12pa}.}
	\label{fig:SI3bis_spatiotemp_normed_12pa}
\end{figure*}

\begin{figure*}
	\centering
    \includegraphics[scale = 0.8]{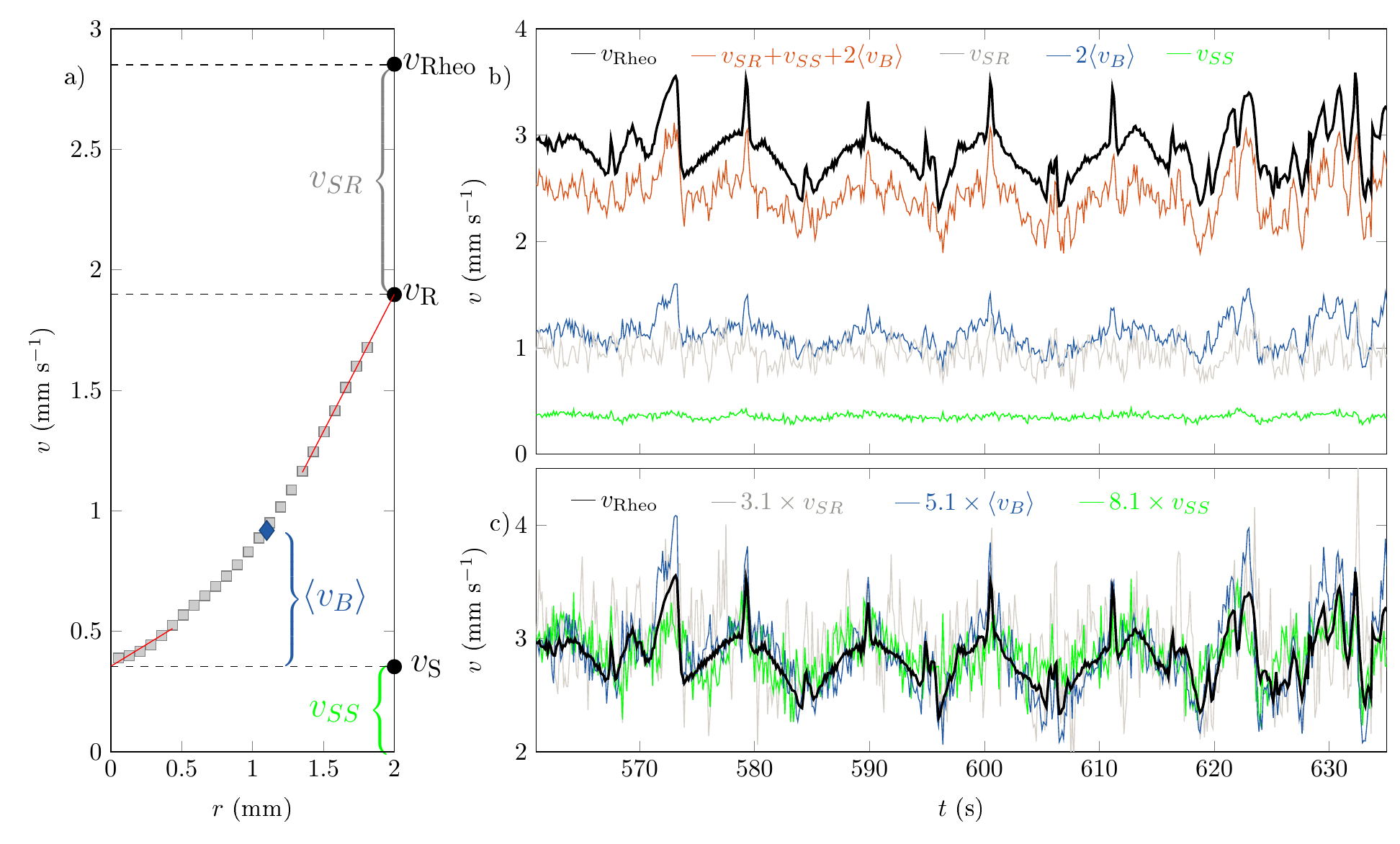}
    \caption{Velocities in regime I for $\sigma = 12$~Pa. (a)~Average velocity profile $\langle v(r,z,t)\rangle_{z,t}$ as a function of the radial distance $r$ to the stator. The red lines show the linear fits used to extrapolate the velocity of the sample at the walls. (b)~Time series of the various velocities defined in Appendix~\ref{s:velo} and highlighted in (a). (c)~The same velocities rescaled so as to match the time average of $v_{\rm Rheo}(t)$.}
	\label{fig:SI2_velocity_12pa}
\end{figure*}

\begin{figure*}
	\centering
    \includegraphics[scale = 0.85]{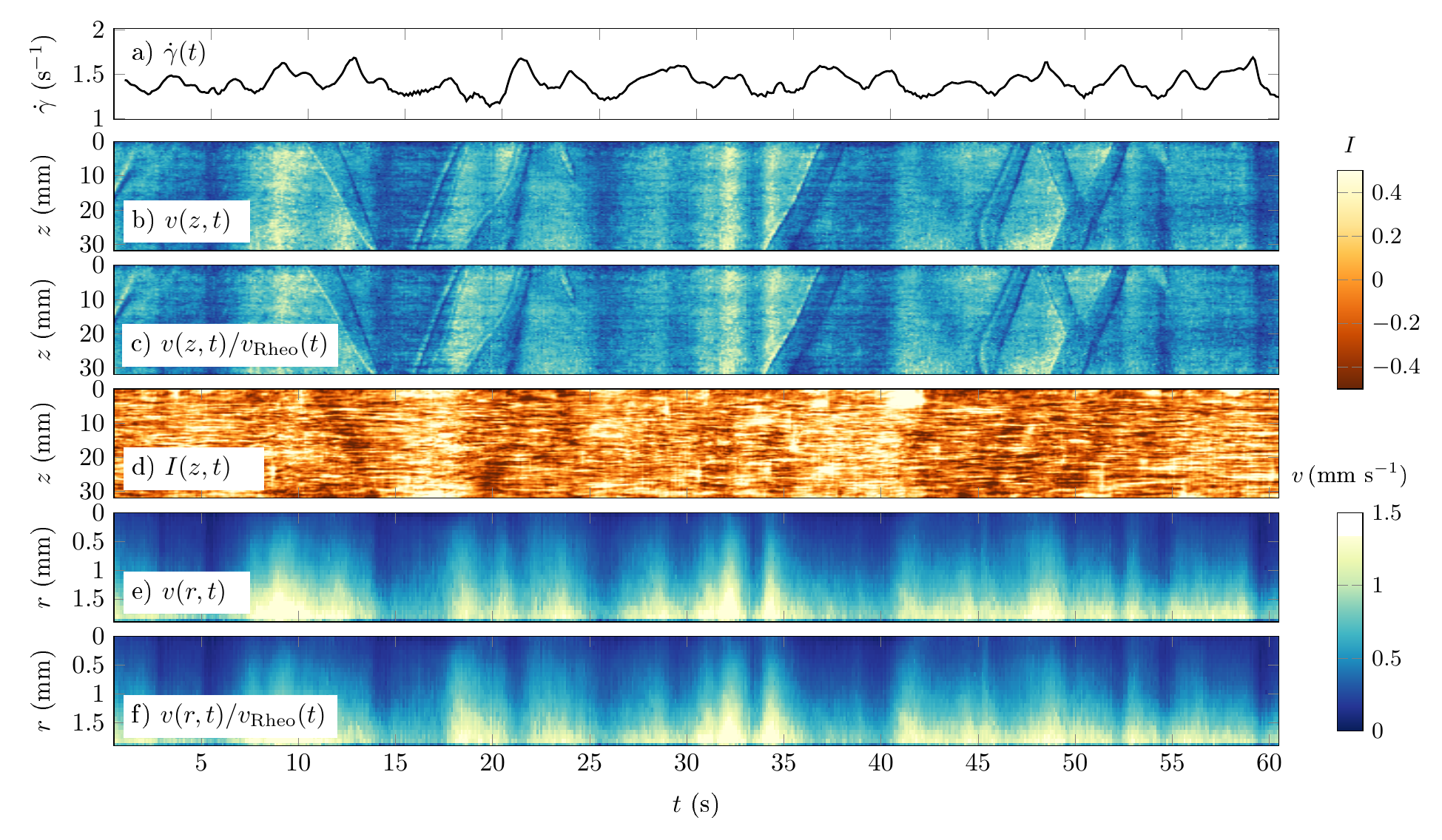}
    \caption{Same as Fig.~\ref{fig:SI3bis_spatiotemp_normed_12pa} in regime II for $\sigma = 80$~Pa. See also Supplemental Video~\ref{vid:80pa}.}
	\label{fig:SI3bis_spatiotemp_normed_80pa}
\end{figure*}

\begin{figure*}
	\centering
    \includegraphics[scale = 0.8]{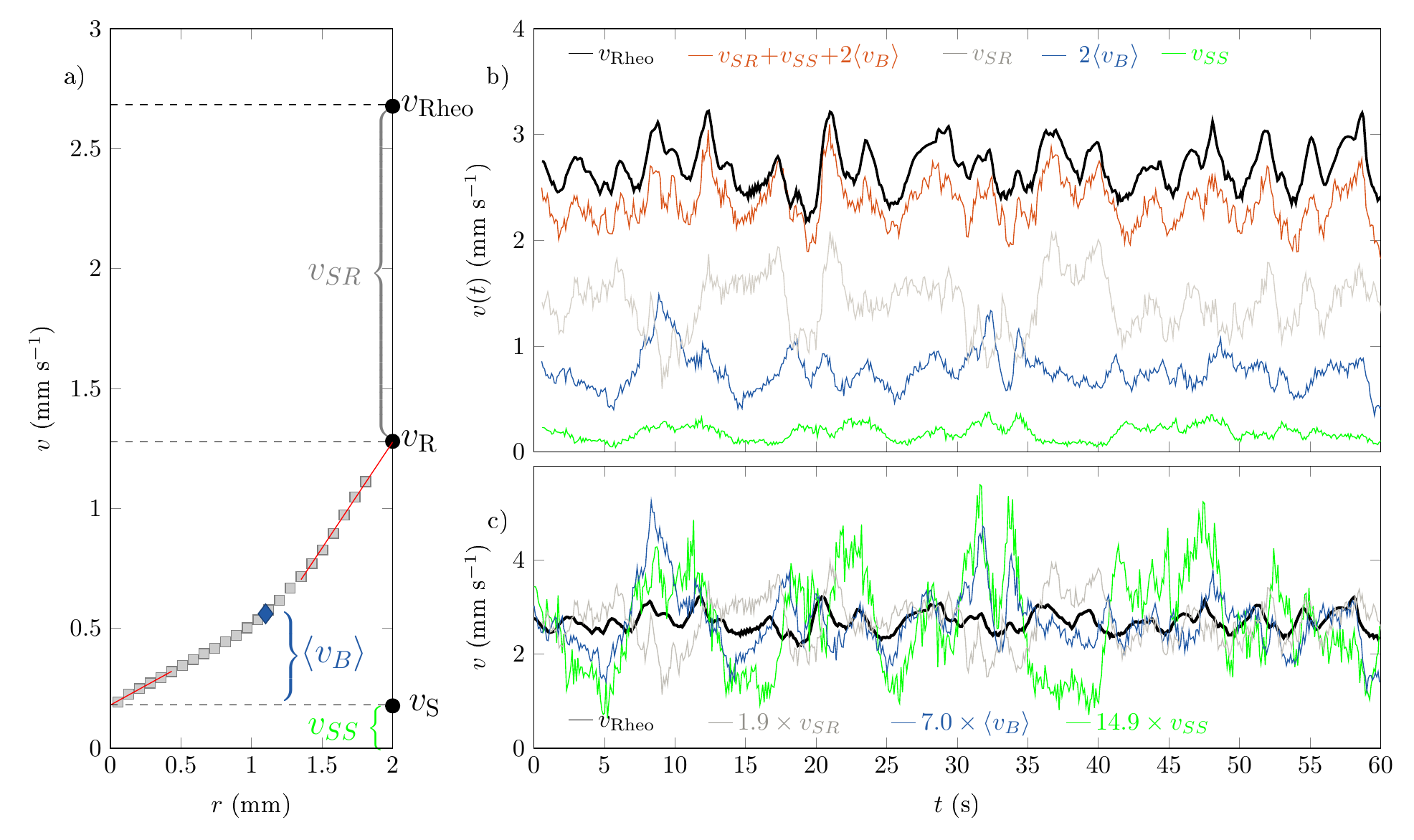}
    \caption{Same as Fig.~\ref{fig:SI2_velocity_12pa} in regime II for $\sigma = 80$~Pa.}
	\label{fig:SI2_velocity_80pa}
\end{figure*}

\clearpage

\subsection{Analysis of local velocity profiles}
\label{s:velo}

Local velocity profiles show significant wall slip both at the rotor and at the stator. In order to compare the dynamics of the local velocity with the global dynamics inferred from the shear rate $\dot\gamma(t)$ measured by the rheometer, we consider the $z$-averaged velocity profiles $v(r,t)$ and estimate the sample velocities $v_{\rm S}(t)$ and $v_{\rm R}(t)$ respectively at the stator and at the rotor through linear extrapolations of $v(r,t)$ (see red lines in Figs.~\ref{fig:SI2_velocity_12pa}a and \ref{fig:SI2_velocity_80pa}a). Slip velocities are then simply given by $v_{SS} (t)= v_{\rm S}(t)$ at the stator and $v_{SR}(t)=v_{\rm Rheo}(t)-v_{\rm R}(t)$ at the rotor, where $v_{\rm Rheo}(t)$ is the velocity of the rotor at time $t$.  $v_{\rm Rheo}$ is linked to the global shear rate through $v_{\rm Rheo}(t) = R_i (R_o^2 - R_i^2)/(R_o^2 + R_i^2) \dot\gamma(t)\simeq \dot\gamma(t) e$, where the last approximation results from the small-gap limit $e\ll R_i$. The ``bulk'' velocity profile is then obtained by subtracting the slip velocity at the stator to $v(r,t)$: $v_B(r,t) = v(r,t) - v_{SS}(t)$. Taking the average over $r$, we get $\langle v_B \rangle(t)$ from which we define the effective shear rate felt by the bulk material as $\dot\gamma_B(t)=(R_o^2 + R_i^2)/[R_i (R_o^2 - R_i^2)]\langle v_B \rangle(t)\simeq \langle v_B \rangle(t)/e$.

Figures~\ref{fig:SI2_velocity_12pa} and \ref{fig:SI2_velocity_80pa} show the results of the analysis of the velocity profiles corresponding to the time sequences of  Figs.~\ref{fig:SI3bis_spatiotemp_normed_12pa} and \ref{fig:SI3bis_spatiotemp_normed_80pa}. In regime I, we observe a very strong positive correlation between $v_{\rm Rheo}$, $v_{SS}$ and $\langle v_B\rangle$ and a fairly positive correlation between $v_{\rm Rheo}$ and $v_{SR}$, see Fig.~\ref{fig:SI2_velocity_12pa}c. In regime II, however, the various velocities show very different dynamics, see Fig.~\ref{fig:SI2_velocity_80pa}b,c. The fact that $v_{SS}+v_{SR}+2\langle v_B\rangle$ does not exactly match $v_{\rm Rheo}$ can be ascribed to the downward curvature of the velocity profiles that leads to an average value $\langle v_B\rangle$ smaller than $(v_R-v_S)/2$. 

\subsection{Flow curves for various concentrations}
\label{s:concentration}

It is well established that the rheological properties of cornstarch not only depend on its concentration but also on its polydispersity, on its porosity, on the ambient humidity and on the supplier \cite{Brown:2014,Han2016,Han:2017}. Moreover different authors have used different suspending fluids, e.g. pure water \cite{fall2015}, water--glycerol mixtures \cite{hermes2016} or water--CsCl mixtures as in the present work \cite{brown2009,Han2016,Han:2017}. Therefore comparisons with previous or future studies require to locate our sample within the shear-thickening phase diagram of cornstarch. Although a systematic investigation of the influence of the concentration on the effects studied here is out of the scope of the present paper, Fig.~\ref{fig:concentration} shows flow curves measured on our cornstarch system at different weight fractions in a parallel-plate geometry. The transition from continuous to discontinuous shear thickening occurs at about 38\%~wt., above which the flow curves clearly show a sigmoidal part. The corresponding volume fraction has been noted $\phi_{\rm c}$ or $\phi_{\rm DST}$ in previous works \cite{wyart2014,fall2015,hermes2016,chacko2018}. Further tests at 43\%~wt. reveal strong surface instability at low shear stress so that such a large weight fraction most probably corresponds to a volume fraction above the jamming point for frictional particles, noted $\phi_{\rm m}$ or $\phi^\mu_{\rm J}$ in Refs.~\cite{wyart2014,hermes2016,chacko2018}. We conclude that the weight fraction of 41\% wt. studied in details in the present paper lies close to full jamming but still belongs to the DST region of the phase diagram.

\subsection{DST in other geometries}
\label{s:geo}

In order to test for the robustness of the dynamics observed in our smooth Taylor-Couette geometry, we conduct additional experiments in a rough Taylor-Couette geometry and in a rough parallel-plate geometry, see Fig.~\ref{fig:geometry}. The rough Taylor-Couette geometry uses a rotor of radius $R_i = 23.9$~mm and height $H=61$~mm covered with sandpaper (P-320 grade, grain size 46~$\mu$m) and a stator of radius $R_o = 25.0$~mm made of sandblasted PMMA (typical roughness of a few micrometers). The gap width of this Taylor-Couette cell is thus $e=1.1$~mm. Strong scattering of ultrasound due to the roughness of the stator makes it impossible to perform ultrasonic imaging simultaneously to rheometry with this Taylor-Couette cell. Consequently we could not check for wall slip or propagating events in this case. The parallel-plate device consists of two plates of diameter $29$~mm covered with the same sandpaper as the rough rotor and separated by a gap of $1$~mm. 

The data in Fig.~\ref{fig:geometry} reveal the same phenomenology as that found with the smooth Taylor-Couette geometry of gap 2~mm. Both flow curves show the distinctive vertical portion characteristic of DST. Within the DST regime, the shear rate measured under an imposed stress becomes unsteady. The critical shear rate and shear stress at DST onset inferred from stress sweeps are slightly less than those in Fig.~\ref{fig:fc_tseries_spectra}a, which could indicate some degree of sensitivity to the geometry and/or sample preparation. The fact that, in the parallel plates, long experiments under constant stress record shear rates that are increasingly larger than in the preceding, rather fast stress ramp is most probably due to humidity absorption by the suspension. As suggested in Ref.~\cite{hermes2016}, particle migration is also more likely to occur in the parallel-plate geometry. In any case, the shear rate $\dot\gamma(t)$ always slowly drifts to larger values such that the shear rate response in parallel plates is not strictly statistically stationary. This precludes a detailed statistical analysis of long signals as described in the main text. Still, the fluctuations observed in Fig.~\ref{fig:geometry} are qualitatively similar to those of Fig.~\ref{fig:fc_tseries_spectra}b. Therefore, they are likely to be due to the same local dynamical phenomena, although this remain to be fully investigated and demonstrated from local measurements.

\begin{figure*}
	\centering
    \includegraphics[scale = 0.85]{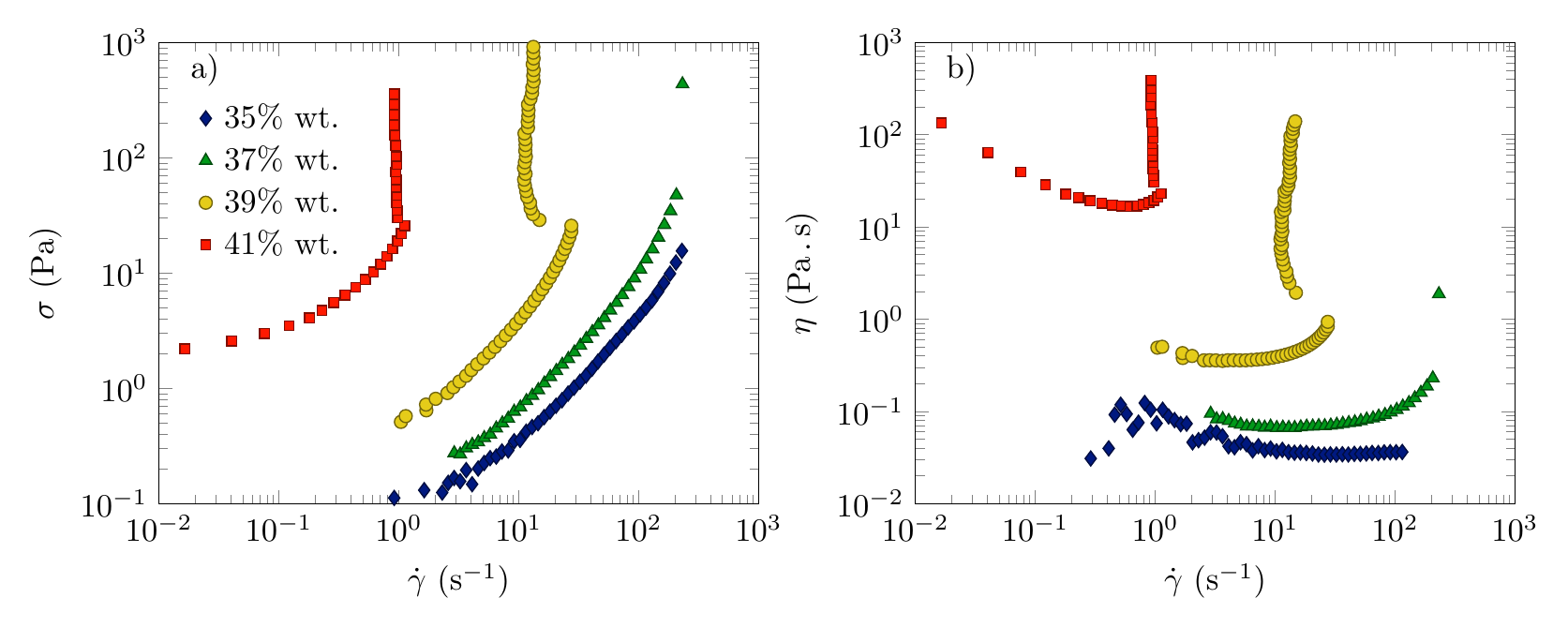}
 	\caption{Flow curves of dense cornstarch suspensions at various weight fractions and prepared as described in Appendix~\ref{s:sample}. (a) Shear stress $\sigma$ and (b) viscosity $\eta$ as a function of the shear rate $\dot\gamma$. Data obtained by ramping up the imposed stress $\sigma$ with at least 5~s per point in the parallel-plate geometry described in Appendix~\ref{s:geo}.}
    \label{fig:concentration}
\end{figure*}

\begin{figure*}
	\centering
    \includegraphics[scale = 0.85]{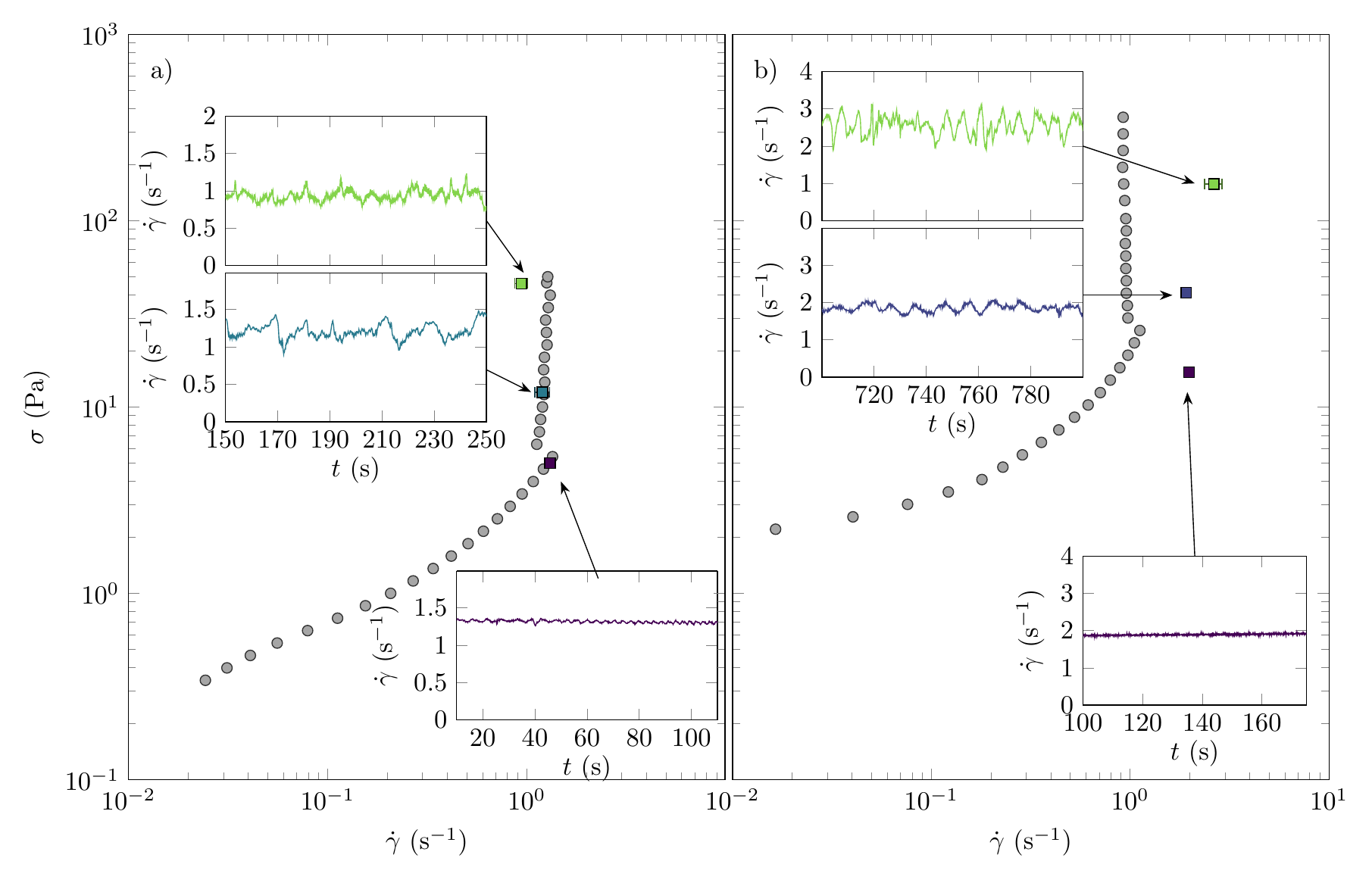}
    \caption{DST in a dense cornstarch suspension at 41\% wt. for two different rough shearing geometries: (a) a small-gap Taylor-Couette cell with a rotor covered with sandpaper and a sandblasted stator and (b) a parallel-plate device with walls covered with sandpaper. The main graphs show the flow curves recorded by ramping up the imposed stress $\sigma$ with 20~s per point in (a) and 10~s per point in (b) (grey circles) and a few subsequent long experiments under constant stresses (colored circles, the shear rate $\dot\gamma$ is averaged over at least 300~s and error bars correspond to the standard deviation). The insets are snapshots of long time series of $\dot\gamma (t)$ for three different imposed stresses across DST.}
	\label{fig:geometry}
\end{figure*}

\clearpage

\section{Supplemental videos}

\begin{video*}[h!]
	\centering
    \includegraphics[scale = .66]{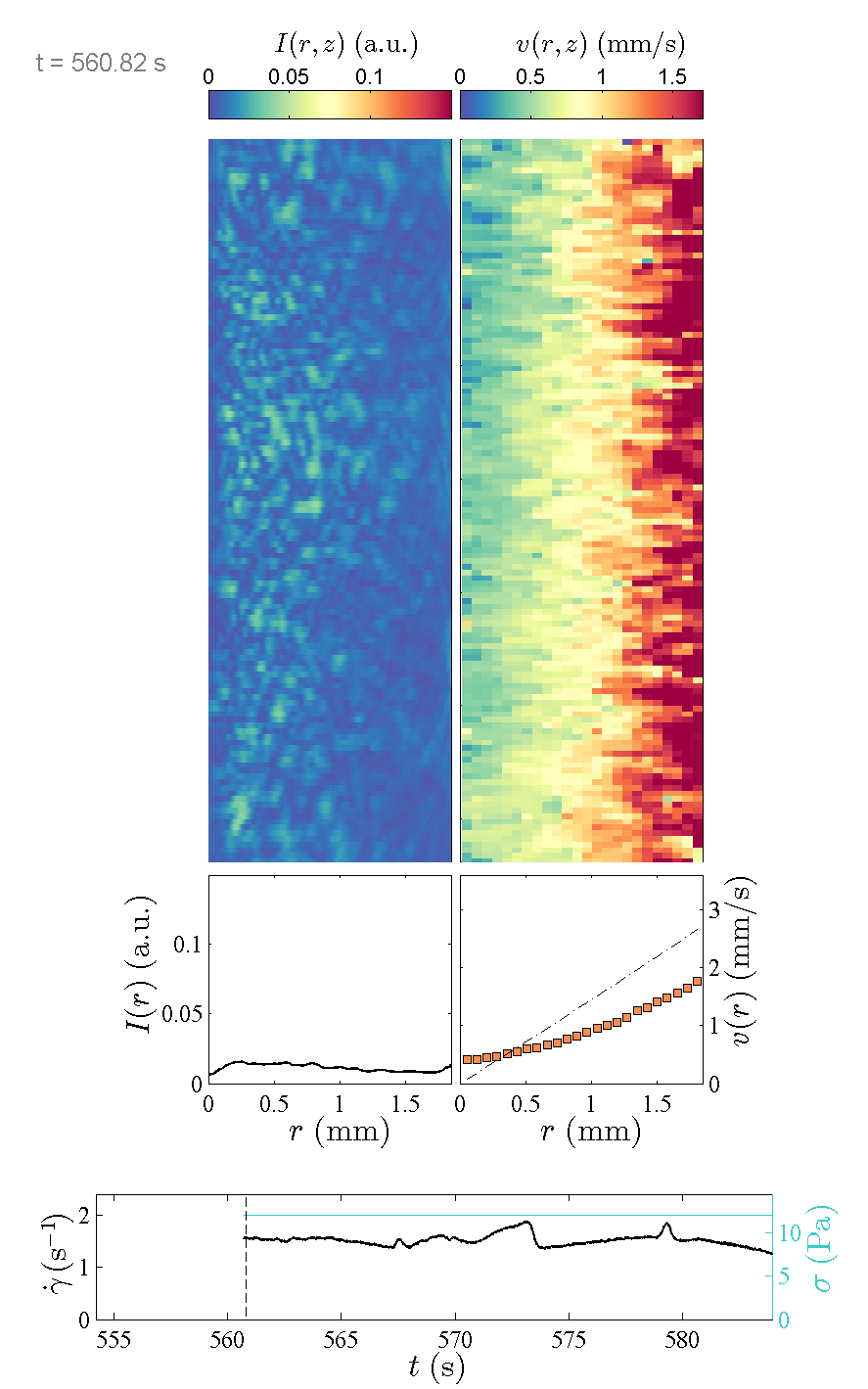}
    \setfloatlink{./Video1_12Pa.mp4}
    \caption{Video of the experiment at $\sigma=12$~Pa (regime I). Top panels: instantaneous maps of the ultrasonic intensity $I(r,z,t)$ (left) and of the tangential velocity $v(r,z,t)$ (right). The vertical $z$-axis covers 32 mm and the horizontal $r$-axis spans the entire gap of 2~mm. The rotor is located at $r=2$~mm. Middle panels: same data averaged over the vertical direction $z$. Bottom panels: rheological signals $\dot\gamma(t)$ (black) and $\sigma(t)$ (blue) recorded by the rheometer simultaneously to the ultrasonic data. The vertical dashed line indicates the current time in the video.}
	\label{vid:12pa}
\end{video*}

\clearpage

\begin{video*}[h!]
	\centering
    \includegraphics[scale = .66]{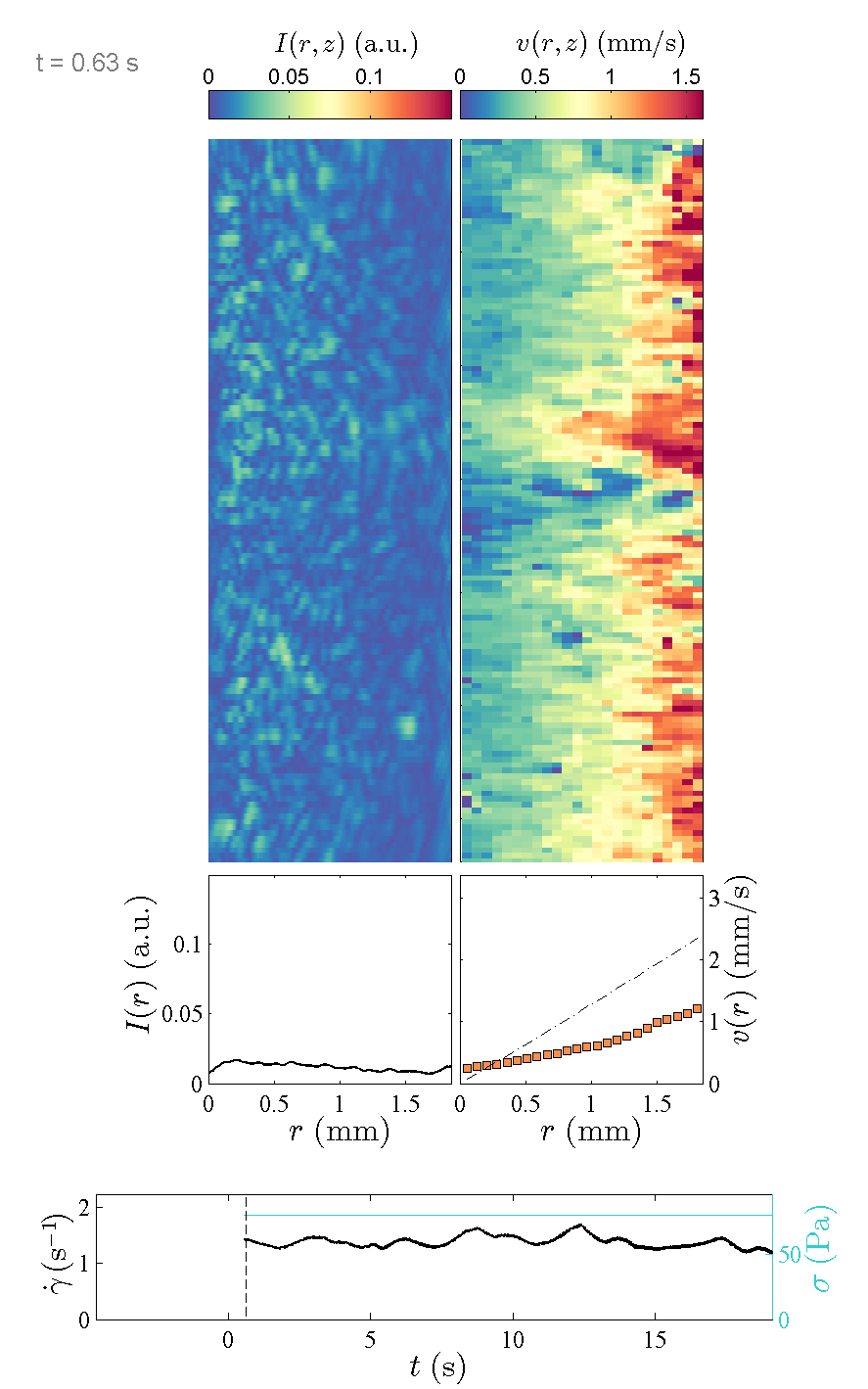}
    \setfloatlink{./Video2_80Pa.mp4}
    \caption{Video of the experiment at $\sigma=80$~Pa (regime II). Same caption as in Video~\ref{vid:12pa}.}
	\label{vid:80pa}
\end{video*}

\end{widetext}

\end{document}